\documentclass[
	reprint,
showkeys,
	aps,
	prd,
	longbibliography,
nofootinbib, 
onecolumn
]{revtex4-2}

\usepackage[utf8]{inputenc}
\usepackage{tabularx,ragged2e}
\newcolumntype{Y}{>{\centering\arraybackslash}X} 
\usepackage{amsmath}
\usepackage{amssymb}
\usepackage{upgreek}
\usepackage{physics}
\usepackage{slashed}

\usepackage{hhline}
\usepackage{dsfont}

\usepackage[hyperref]{xcolor}
	\definecolor{goethe-blau}{cmyk}{1.0,0.2,0.0,0.4}
	\definecolor{hellgrau}{cmyk}{0.04,0.04,0.05,0.02}
	\definecolor{sandgrau}{cmyk}{0.12,0.09,0.13,0.0}
	\definecolor{dunkelgrau}{cmyk}{0.25,0.25,0.30,0.75}
	\definecolor{emo-rot}{cmyk}{0.04,1.0,0.8,0.07}
	\definecolor{purple}{cmyk}{0.08,1.0,0.3,0.36}
	\definecolor{senfgelb}{cmyk}{0.01,0.25,1.0,0.05}
	\definecolor{gruen}{cmyk}{0.62,0.4,0.87,0.09}
	\definecolor{magenta}{cmyk}{0.08,0.86,0.12,0.12}
	\definecolor{orange}{cmyk}{0.0,0.7,1.0,0.04}
	\definecolor{sonnengelb}{cmyk}{0.0,0.12,0.95,0.0}
	\definecolor{helles-gruen}{cmyk}{0.4,0.17,0.81,0.07}
	\definecolor{lichtblau}{cmyk}{0.8,0.0,0.06,0.04}

\usepackage[
	colorlinks,
	pdfpagelabels,
	breaklinks,
	pdfstartview=FitH,
	bookmarksopen=true,
	bookmarksnumbered=true,
	bookmarksopenlevel=2,
	plainpages=false,
	hypertexnames=false,
	citecolor=emo-rot,
	linkcolor=goethe-blau,
	urlcolor=purple,
	pdftitle={Absence of inhomogeneous chiral phases in 2+1-dimensional four-fermion and Yukawa models}, 
	pdfauthor={Laurin Pannullo, Marc Winstel},
	pdfkeywords={}
]{hyperref}

\usepackage[capitalise]{cleveref}
\usepackage{multirow}
\usepackage{xstring}
\usepackage{graphicx}	
\usepackage[acronym]{glossaries-extra}
\setabbreviationstyle[acronym]{long-short}

\newacronym{ff}{FF}{four-fermion}
\newacronym[plural=QFTs, longplural=quantum field theories]{qft}{QFT}{quantum field theory}
\newacronym{qcd}{QCD}{Quantum Chromodynamics}
\newacronym{lp}{LP}{Lifshitz point}
\newacronym{njl}{NJL}{Nambu-Jona-Lasinio}
\newacronym{cnjl}{PSFF}{complete Lorentz-(pseudo)scalar four-fermion}
\newacronym{qm}{QM}{quark-meson}
\newacronym{gn}{GN}{Gross-Neveu}
\newacronym{chign}{$\chi$HGN}{chiral Heisenberg-Gross-Neveu}
\newacronym{ip}{IP}{inhomogeneous phase}

\glssetcategoryattribute{acronym}{nohyperfirst}{true}

\providecommand{\Rcite}[1]{\begingroup
	\def\tempx{0}\StrCount{#1}{,}[\tempx]\ifnum\tempx > 0 
	Refs.~\else
	Ref.~\fi
	\endgroup
	\cite{#1}}

\newcommand{\infN}{\ensuremath{\N \to \infty}}
\newcommand{\Q}{\mathrm{Q}}
\newcommand{\D}{\Q}
\newcommand{\ii}{\ensuremath{\mathrm{i}}}
\newcommand{\e}{\ensuremath{\mathrm{e}}}
\newcommand{\I}{\ensuremath{\mathds{I}}}
\newcommand{\U}{\ensuremath{\mathrm{U}}}
\newcommand{\SU}{\ensuremath{\mathrm{SU}}}
\newcommand{\Det}{\ensuremath{\mathrm{Det}}}

\newcommand{\stV}[1]{\ensuremath{#1}}
\newcommand{\xstV}{\ensuremath{\stV{x}}}

\newcommand{\sV}[1]{\ensuremath{\mathbf{#1}}}
\newcommand{\xsV}{\ensuremath{\sV{x}}}
\newcommand{\qsV}{\ensuremath{\sV{q}}}

\newcommand{\Hom}[1]{\ensuremath{\bar{#1}}}
\newcommand{\HD}{\Hom{\D}}

\newcommand{\auxf}{\ensuremath{\phi}}
\newcommand{\dbf}{\ensuremath{\varphi}}
\newcommand{\Yf}{\ensuremath{\chi}}
\newcommand{\dbYf}{\ensuremath{\zeta}}

\newcommand{\minauxf}{\ensuremath{\Phi}}

\newcommand{\ve}[1]{\ensuremath{\vec{#1}}}
\newcommand{\vauxf}{\ensuremath{\ve{\auxf}}}
\newcommand{\vdbf}{\ensuremath{\ve{\dbf}}}
\newcommand{\vYf}{\ensuremath{\ve{\Yf}}}

\newcommand{\vminauxf}{\ensuremath{\ve{\minauxf}}}

\newcommand{\Hauxf}{\ensuremath{\Hom{\auxf}}}
\newcommand{\Hminauxf}{\ensuremath{\Hom{\minauxf}}}
\newcommand{\vHauxf}{\ensuremath{\ve{\Hauxf}}}
\newcommand{\vminHauxf}{\ensuremath{\ve{\Hminauxf}}}

\newcommand{\HYf}{\ensuremath{\Hom{\Yf}}}
\newcommand{\vHYf}{\ensuremath{\ve{\HYf}}}
\newcommand{\HdbYf}{\ensuremath{\Hom{\dbYf}}}
\newcommand{\vHdbYf}{\ensuremath{\ve{\HdbYf}}}
\newcommand{\Mass}{\ensuremath{M}}
\newcommand{\minMass}{\ensuremath{\mathcal{M}}}

\newcommand{\pt}[1]{\ensuremath{\delta #1}}
\newcommand{\ptauxf}{\ensuremath{\pt{\auxf}}}
\newcommand{\ptvauxf}{\ensuremath{\pt{\vauxf}}}
\newcommand{\ptD}{\ensuremath{\Delta \D}}
\newcommand{\ptdbf}{\ensuremath{\pt{\dbf}}}

\newcommand{\ptYf}{\ensuremath{\pt{\Yf}}}

\newcommand{\ft}[1]{\ensuremath{\tilde{#1}}}
\newcommand{\ptftauxf}{\ensuremath{\pt{\ft{\auxf}}}}

\newcommand{\ptftdbf}{\ensuremath{\pt{\ft{\dbf}}}}
\newcommand{\ftHD}{\ensuremath{\ft{\HD}}}

\newcommand{\ptftdbYf}{\ensuremath{\pt{\ft{\dbYf}}}}

\newcommand{\pauli}{\ensuremath{\tau}}
\newcommand{\vpauli}{\ensuremath{\vec{\pauli}}}

\newcommand{\cm}{\ensuremath{c}}
\newcommand{\intchannelSet}{\ensuremath{C}}
\newcommand{\cmstar}{\ensuremath{c^\star}}
\newcommand{\xnod}{\ensuremath{\tau}}

\newcommand{\dr}{\ensuremath{\mathrm{d}}}
\newcommand{\coupling}{\ensuremath{\lambda}}
\newcommand{\indSet}{\ensuremath{J}}

\renewcommand{\S}{\mathcal{S}}
\newcommand{\seff}{\mathcal{S}_{\text{eff}}}
\newcommand{\Z}{\ensuremath{\mathbb{Z}}}
\newcommand{\N}{\ensuremath{N}}

\newcommand{\gtwo}{\ensuremath{\Gamma^{(2)}}}
\newcommand{\twopid}[1]{(2 \uppi)^{#1}}
\newcommand{\intMeasureOverPi}[2]{\tfrac{\dr^{#1}{#2}}{\twopid{#1}}}
\newcommand{\Nbar}{8}

\newcommand{\matsubarasum}{\ensuremath{\sum_{n=-\infty}^{\infty}}}
\newcommand{\piveccomp}[2]{\ensuremath{\pi_{#1, #2}}}

\newcommand{\ndist}[2]{n(\beta(#1+#2))}
\newcommand{\nbardist}[2]{n(\beta(#1-#2))}
\newcommand{\Ell}{L}
\newcommand{\EllOne}{\Ell_1}

\newcommand{\EllOneArgs}[3]{\EllOne{}(#1,#2,#3)}
\newcommand{\EllTwo}[1]{\Ell_{2,#1}}
\newcommand{\EllTwoPlus}{\EllTwo{+}}
\newcommand{\EllTwoMinus}{\EllTwo{-}}
\newcommand{\EllTwoPlusMinus}{\EllTwo{\pm}}
\newcommand{\EllTwoArgs}[5]{\EllTwo{#1}(#2,#3,#4,#5)}

\newcommand{\mycell}[1]{%
	\begin{tabular}[t]{@{}l@{}} #1 \end{tabular}}

\newcommand{\mycellcenter}[1]{%
	\begin{tabular}[t]{@{}c@{}} #1 \end{tabular}}

\begin{document}

\preprint{APS/123-QED}

\title{
	Absence of inhomogeneous chiral phases in \texorpdfstring{$(2 + 1)$}{(2+1)}-dimensional four-fermion and Yukawa models
}

\author{Laurin Pannullo}
	\email{pannullo@itp.uni-frankfurt.de}

\author{Marc Winstel}
	\email{winstel@itp.uni-frankfurt.de}
		
	\affiliation{
		Institut für Theoretische Physik,  Goethe-Universität Frankfurt am Main,
		\\
		Max-von-Laue-Straße 1, D-60438 Frankfurt am Main, Germany.
	}
	
\date{\today}

\begin{abstract}
	We show the absence of an instability of homogeneous (chiral) condensates against spatially inhomogeneous perturbations for various $2+1$-dimensional four-fermion and Yukawa models.
	All models are studied at non-zero baryon chemical potential, while some of them are also subjected to chiral and isospin chemical potential.
	The considered theories contain up to 16 Lorentz-(pseudo)scalar fermionic interaction channels.
	We prove the stability of homogeneous condensates by analyzing the bosonic two-point function, which can be expressed in a purely analytical form at zero temperature.
	Our analysis is presented in a general manner for all of the different discussed models.
	We argue that the absence of an inhomogeneous chiral phase (where the chiral condensate is spatially non-uniform) follows from this lack of instability.
	Furthermore, the existence of a moat regime, where the bosonic wave function renormalization is negative, in these models is ruled out.		

\end{abstract} 

\maketitle

\tableofcontents

\section{Introduction\label{sec:intro}}
	\Glspl{qft} with \gls{ff} interactions, so-called \gls{ff} models, are relevant for several branches of physics and, recently, much attention has been paid to their investigation.
	\gls{ff} models, such as the \gls{njl} model \cite{PhysRev.122.345, PhysRev.124.246}, and also Yukawa-type models, such as the \gls{qm} model \cite{Gell-Mann:1960mvl}, are considered as low-energy effective models of \gls{qcd} and light meson physics, describing, e.g., spontaneous chiral symmetry breaking or color superconductivity.
	The models are applied to study strongly-interacting, fermionic matter under (extreme) external conditions such as, e.g., temperature, density, chiral or isospin imbalance or magnetic fields (see, e.g., \Rcite{Asakawa:1989bq,Scavenius:2000qd,Hands:2002mr,Sadzikowski:2002iy,Buballa:2003qv, Hands:2004uv, Schaefer:2004en, Sadzikowski:2006jq, Schaefer:2006ds,Hiller:2008eh, Gatto:2012sp, Andersen:2014xxa, Farias:2016let,Khunjua:2018sro,  Braguta:2019pxt, Khunjua:2021pfa, Lopes:2021tro, Ayala:2023cnt}).
	In contrast to \gls{qcd}, the models do not exhibit a sign problem for finite baryon chemical potential rendering, e.g., lattice investigations possible (see, e.g., \Rcite{Walters:2004ya,Hands:2004uv,Pannullo:2022eqh}). 
	
	Often, \gls{ff} models are studied in two spatial dimensions, where they are, in contrast to three spatial dimensions, renormalizable \cite{Rosenstein:1988pt, Gat:1991bf}.
	The motivation of many works \cite{Klimenko:1987gi, Hands:1992be,Hands:1992ck, Inagaki:1994ec, DelDebbio:1995zc,DelDebbio:1997dv, Hands:1999id, Appelquist:2000mb, Hands:2000gv, Hands:2001cs, Allton:2002mv, Hands:2003dh,Strouthos:2003js,zhukovsky:2017hzo, Mandl:2022ryj, Lenz:2023wvk} is the study of physical phenomena in high energy physics such as chiral symmetry breaking or color-superconductivity in rather simple, low-dimensional \glspl{qft}. 
	For example, the phenomenon of spontaneous chiral symmetry breaking induced by an external magnetic field has first been observed in a $2+1$-dimensional \gls{gn}-type model \cite{Klimenko:1990rh, Klimenko:1991he} (see \Rcite{Gross:1974jv} for the original definition of the \gls{gn} model in $1+1$ dimensions). 
	Also, $2+1$-dimensional \gls{ff} and Yukawa models are interesting for the study of technical aspects and the development of techniques in \gls{qft}, see \Rcite{Rosenstein:1988pt, Klimenko:1990my,  Gat:1991bf, Hands:1992be, Klimenko:1993iz,  Kneur:2007vj, Braun:2010tt, Lenz:2019qwu,Hands:2020itv,   Hands:2021eyc,  Hands:2021mrg,  Bonati:2022fiu,  Gubaeva:2022feb}. 
	Another motivation for the study of these models is the application to condensed matter systems with an effectively planar structure, which can be described by a relativistic fermionic dispersion relation\footnote{For example, it was shown that the $1+1$-dimensional \gls{gn} model is suitable for the description of polyacetylene and similar systems \cite{Chodos:1993mf, Thies:2006ti}. In a similar way, $2+1$-dimensional \gls{ff} and Yukawa models can be applied in descriptions of condensed matter systems in $2+1$ dimensions \cite{ Semenoff:1998bk,  Hands:2015lza, Ebert:2015hva, Ebert:2017udh,  Knorr:2017yze,  Gracey:2018cff, Gracey:2018qba, Lang:2018csk}.}. 
	For example, symmetry breaking schemes of graphene effective field theory \cite{Drut:2008rg,Boyda:2013rra, Braguta:2013klm,DeTar:2016dmj,DeTar:2016dmj} are also present in \gls{ff} models.
	Thus, studies  of fermionic models in $2+1$ dimensions under the influence of external parameters, as, e.g., \Rcite{Nunes:2010pv, Klimenko:2012qi, Ebert:2015vua, Ebert:2015hva,ebert:2016ygm, Ebert:2017udh, Kanazawa:2020rsb,Gomes:2022oyo, Mandl:2022ryj, Lenz:2023wvk}, are motivated both from the high energy and the condensed matter perspective. 
	
	In many of the above works the (chiral) order parameters are considered to be homogeneous in space.
	This is a reasonable choice for first investigations of the phase diagram.
	In strongly-interacting systems in condensed matter physics, however, crystalline-like ground states are quite common \cite{Fulde:1964zz, Larkin:1964wok, mertschingIncommensuratePeierlsPhase1981,Gruner:1994zz, Bulgac:2008tm,Radzihovsky:2011zz, Roscher:2013cma, Attanasio:2021jkk} suggesting that the assumption of homogeneous order parameters has to be reevaluated.  
	More refined investigations of \gls{ff} and Yukawa-type models have indeed revealed the presence of a chiral \gls{ip}, where the corresponding order parameter is a function of the spatial coordinates, also in effective theories in nuclear and high energy physics, see \Rcite{Dautry:1979bk, Kutschera:1990xk, Kutschera:1991rh}, where \glspl{ip} have first been found in this context.
	This has stimulated conjectures that inhomogeneous chiral condensates and related phenomena might also be relevant in the phase diagram of \gls{qcd}. 
	However, it has to be remarked that many of the existing model investigations have been carried out within the mean-field approximation, where the bosonic quantum fluctuations are neglected.
	
	One of the most prominent examples for an \gls{ip} is the chiral kink in the $1+1$-dimensional \gls{gn} model\footnote{In condensed matter, the equivalent model is known as the Fröhlich model \cite{mertschingIncommensuratePeierlsPhase1981}.} \cite{Thies:2003kk,Thies:2006ti}, but also various other types of \glspl{ip} with a more complicated structure have been found in $1+1$-dimensional models \cite{Schon:2000he,Basar:2009fg,Thies:2018qgx,Thies:2019ejd,Thies:2020ofv,Thies:2021shf,Thies:2022kuv}.
	In recent literature, there is an on-going discussion whether these phases persist when allowing for bosonic quantum fluctuations \cite{Lenz:2020bxk, Lenz:2021kzo, Lenz:2021vdz, Nonaka:2021pwm,Stoll:2021ori, Ciccone:2022zkg}, where a spontaneous breaking of translational invariance, as in an \gls{ip}, should be forbidden in $1+1$ dimensions according to the Mermin-Wagner theorem \cite{Mermin:1966fe,Coleman:1973ci,Witten:1978qu,Watanabe:2019xul}. 
	
	In $3+1$ dimensions, many of the models for spontaneous chiral symmetry breaking in \gls{qcd} feature an \gls{ip} -- typically appearing at low temperature and intermediate densities, where one would expect a first-order phase transition between homogeneous phases, see \Rcite{Buballa:2014tba} for a review.
	However, these results are mostly obtained in the mean-field approximation and their predictive power for \gls{qcd} can be questioned.
	In case of the \gls{ff} models, we recently documented a regulator dependence of the \gls{ip} in the \gls{njl} model \cite{Pannullo:2022eqh}, where the existence and shape of the \gls{ip} depends on the chosen regularization scheme. 
	In combination with the non-renormalizability of the \gls{njl} model this raises questions about the predictive power for \gls{qcd}, especially since the chemical potentials in the region of the \gls{ip} are in the order of the necessary regulator. 
	The results in the renormalizable \gls{qm} model suffer from other technical problems, such as instabilities at large field values \cite{Carignano:2014jla, Carignano:2016jnw}.
	
	Besides these difficulties of the model studies, there are still indications that inhomogeneous chiral condensates might be relevant in \gls{qcd} \cite{Deryagin:1992rw, Muller:2013tya}. 
	A recent Functional Renormalization Group study of \gls{qcd} \cite{Fu:2019hdw} finds a so-called moat or Lifshitz regime, where the bosonic wave-function renormalization is negative and a modified dispersion relation is obtained \cite{Pisarski:2021qof}. 
	Such a phenomenon is often related to an \gls{ip} \cite{Koenigstein:2021llr}.
	Experimental signals of this regime are discussed in \Rcite{Pisarski:2020dnx, Pisarski:2020gkx, Pisarski:2021qof, Rennecke:2023xhc}. 
	
	In $2+1$-dimensional models, only recently the attention has moved towards the study of \glspl{ip} \cite{Winstel:2019zfn, Narayanan:2020uqt, Buballa:2020nsi,  Winstel:2021yok, Pannullo:2021edr, Winstel:2022jkk, Lenz:2023wvk}.
	In the $2+1$-dimensional \gls{gn} model, \glspl{ip} were observed at finite lattice spacing \cite{Winstel:2019zfn}, but it turned out that these are regulator artifacts depending on the chosen regularization scheme \cite{Narayanan:2020uqt, Buballa:2020nsi}.
	After renormalization homogeneous phases are favored. 
	This regulator dependence further highlights the problem of non-renormalizability with respect to high chemical potentials relevant in the \gls{ip} of the $3+1$-dimensional \gls{njl} model \cite{Pannullo:2022eqh}. 
	In lattice simulations of the $2+1$-dimensional \gls{gn} model, oscillating, but damped correlators have been observed in \Rcite{Hands:2003dh, Strouthos:2003js}, but considering the found regulator dependence of the \gls{ip}  \cite{Narayanan:2020uqt, Buballa:2020nsi} it is unclear whether these are reminiscent from the above described regulator artifacts or from underlying physical reasons, such as Fermi surface effects. 
	We note that the existence of an \gls{ip} is also not favored by the introduction of chiral imbalance \cite{Pannullo:2021edr, Winstel:2021yok} nor of a magnetic field \cite{Lenz:2023wvk}.
	
	In this paper, we rule out the instability of homogeneous condensates with respect to inhomogeneous perturbations for a large class of models with Lorentz-(pseudo)scalar isospin \gls{ff} interactions.
	This extends the previous findings in the \gls{gn} model to a variety of different \gls{ff} and Yukawa models.
	We apply the framework of analyzing the stability of the bosonic two-point functions (as, e.g., used in \Rcite{Nakano:2004cd, deForcrand:2006zz, Wagner:2007he, Tripolt:2017zgc, Buballa:2018hux, Winstel:2019zfn, Buballa:2020xaa, Buballa:2020nsi, Koenigstein:2021llr, Pannullo:2021edr, Pannullo:2022eqh, Winstel:2022jkk}) to these models.
	This analysis tests whether homogeneous field configurations are energetically unstable against inhomogeneous perturbations.
	Among the limitations of the method is that one can only show the existence of an \gls{ip} and not the shape of energetically preferred inhomogeneous field configuration (see \Rcite{Koenigstein:2021llr} for a detailed discussion of this method).
	As discussed in \Rcite{Koenigstein:2021llr}, there might still exist an \gls{ip}, which is not detected by the stability analysis. 
	However, to our knowledge no phase diagram has been found, where an \gls{ip} does not at least enclose a region in parameter space with instabilities of the homogeneous minimum against inhomogeneous perturbations.
	Moreover, all \glspl{ip} so-far observed in model studies are connected by a second-order phase transition to the chirally symmetric phase. 
	Such a phase transition can always be detected by analyzing the stability of the symmetric minimum of the effective potential.
	
	This work is outlined as follows.
	A general \gls{ff} model, containing all the studied interaction channels, is introduced in \cref{sec:theory}.
	Furthermore, we present an extension of these models to Yukawa models.
	In \cref{sec:stability}, the stability analysis of the bosonic two-point function for finite baryon chemical potential and temperature is presented for the \gls{ff} models and their extension to Yukawa models.
	Based on this analysis, we argue that all models with Lorentz-(pseudo)scalar interaction channels do not exhibit \glspl{ip} or moat regimes.
	In \cref{sec:results}, examples for these \gls{ff} models are presented.
	Allowing for multiple chemical potentials, we also show the stability of homogeneous condensates for a few theories each containing a small subset of the previously discussed interaction channels.
	Finally, we sum up our results and conclude in \cref{sec:conclusion_and_outlook}.

\section{Definition of the considered models}
	\label{sec:theory}
	In this section, we introduce a \gls{ff} model, which serves as the general prototype for the models, which will be later studied using the stability analysis introduced in \cref{sec:stability}. 
	We perform a bosonization with auxiliary fields and obtain the effective action. 
	Finally, in \cref{sec:I_Yukawa}, the Yukawa models, obtained from extending the bosonized \gls{ff} models, are defined. 
\subsection{Four-fermion models}
	In order to set up our general analysis, we define
	\begin{align}
		&\mathcal{S}_{\mathrm{FF}}[\bar{\psi},\psi]   \label{eq:FFmodel}   
		= \int_0^\beta \! \dr \tau \int\! \dr^2 \xstV   \left\{ \bar{\psi}\left(\slashed{\partial}+ \gamma_3 \mu \right)  \psi -  \left[ \sum_{j=1}^{16} \tfrac{\coupling_j}{2 \N } \left(\bar{\psi}\,  \cm_j\, \psi\right)^2  \right]\right\}, 
	\end{align} 
	as the most general \gls{ff} action studied in this work in $2+1$-dimensional Euclidean spacetime ($\xstV = (\xsV, \xnod)$ represents the spacetime coordinate).
	The integration over the periodic Euclidean temporal coordinate $\xnod$ goes from $0$ to $\beta = 1 / T$, where $T$ is the temperature, while the integration over $\dr^2 x$ goes over the two-dimensional spatial plane.
	The vector $\psi$ contains $2\N$ four-component fermion fields ($\N$ identical spinors with isospin up/down respectively). We work in the chiral limit, so no bare mass term is introduced. The Dirac matrices are reducible, $4\times4$ representations of the $2+1$-dimensional Euclidean Clifford algebra
	\begin{align}
		\{\gamma_\mu,\, \gamma_\nu\} = \gamma_\mu  \gamma_\nu+ \gamma_\nu \gamma_\mu = 2 \delta_{\mu\nu}\I, \quad \mu, \nu = 1,2,3,\label{eq:clifford}
	\end{align} 
	where $\I$ is the $4\times4$ identity matrix. Useful representations for computations in $2+1$ dimensions can, e.g., be found in \Rcite{Pisarski:1984dj, Gies:2009da, Buballa:2020nsi,  Pannullo:2021edr}.
	The interaction vertices $c_j$ are $8\times8$ matrices in isospin and spin space and elements of  
	\begin{equation}
		\intchannelSet = \left(\cm_j\right)_{j=1,\ldots, 16} = \left(1, \ii\gamma_4, \ii\gamma_5, \gamma_{45}, \vpauli, \ii\vpauli\gamma_4, \ii\vec{\tau}\gamma_5, \vpauli\gamma_{45}\right), \label{eq:dirac_basis}
	\end{equation}
	where $\vpauli$ is the vector of Pauli-matrices acting on the isospin degrees of freedom. The spin matrices $\gamma_4$ and $\gamma_5$ anti-commute with the $\gamma_\nu$, while $\gamma_{45} \equiv \ii \gamma_{4} \gamma_{5}$ commutes with the $\gamma_\nu$.
	All elements of $\intchannelSet$ are $8\times8$ matrices, where the identity matrices in spin and isospin space are not explicitly written down whenever the matrix $\cm_j$ is diagonal in the corresponding space. The channels $\left(\bar{\psi}(\xstV)\,  \cm_j\, \psi(\xstV)\right)^2$ are local, Lorentz-(pseudo)scalar \gls{ff} interactions in the $\SU(2)$ isospin space and spinor space. 
	The vector and matrix-like \gls{ff} channels are not taken into account, since the analysis of these interactions differs technically from the (pseudo-)scalar ones.
	Therefore, these interaction terms will be left to forthcoming work.
	The couplings $\coupling_j$ of each channel have inverse energy dimension and will be set to either $0$ or $\lambda$ in order to study different models and allow for different symmetry groups of the action\footnote{Typically, a FF model, which is invariant under a continuous chiral symmetry transformation, involves two or more FF channels, whose couplings all take an identical value in order to allow for a rotational symmetry transformation between them.}. 
	The chemical potential $\mu$ is introduced in the usual way and induces a non-vanishing baryon density $\propto \bar{\psi} \gamma^3 \psi$. 
	
	It is well known, that the partition function of \cref{eq:FFmodel} is identical to a  partition function with auxiliary bosonic scalar fields $\vauxf$, where $\auxf_j$ can be introduced via a Hubbard-Stratonovich transformation (an inverse shifted Gaussian integration) in order to get rid of the FF interaction $\left(\bar{\psi}(\xstV)\,  \cm_j\, \psi(\xstV)\right)^2$, up to a physically irrelevant integration constant \cite{Gross:1974jv}.
	The partially bosonized action is then given by
	\begin{align}
	    \S[\bar{\psi}, \psi, \vauxf] &= \int \mathrm{d}^3x  \left[\N \sum_{j \in \indSet} \frac{\auxf_j^2 }{2 \lambda_j} + \bar{\psi} \D \psi \right], \quad
	     \D = \slashed{\partial}+ \gamma_3 \mu + \sum_{j \in \indSet} \, \cm_j \, \auxf_j , \label{eq:part_bosonized}
	\end{align}
	where $\indSet$ is an index set containing\footnote{The definition of this set ensures that an auxiliary field $\auxf_k$ is only introduced, when the coupling $\lambda_k$ of the corresponding channel is non-vanishing.} all integers $1 \leq j \leq 16$ with $\lambda_j \neq 0$. 
	The auxiliary fields $\auxf_j$ have the dimension of an energy and fulfill the Ward identity\footnote{The relations for these $1$-point functions can derived imposing the invariance of the functional integration measure $\prod_i \mathcal{D} \auxf_i$ under the infinitesimal shifts of the fields $\vauxf$.}
	\begin{equation}
		\langle \auxf_j\rangle = - \tfrac{\coupling_j}{\N} \langle \bar{\psi} \cm_j \psi\rangle, \quad j \in \indSet. \label{eq:Ward}
	\end{equation}
	In order to relate to the literature about FF models as low-energy effective models (e.g.\ \Rcite{Gell-Mann:1960mvl, PhysRev.122.345, PhysRev.124.246, Scavenius:2000qd,  Buballa:2003qv, Schaefer:2004en, Schaefer:2006ds, Buballa:2014tba, Khunjua:2021pfa, Ayala:2023cnt}), we define phenomenologically motivated symbols for the $\phi_j$, i.e., the tuple
	\begin{equation}
		\Phi = \left(\auxf_j\right)_{j=1, \ldots, 16} = \left(\sigma, \eta_4, \eta_5, \eta_{45}, \ve{a}_0, \ve{\pi}_4, \ve{\pi}_5, \ve{\pi}_{45}\right). \label{eq:phi_sym}
	\end{equation}
	The ordering of this tuple is used to directly map a field $\auxf_j$ to the corresponding fermion bilinear $\bar{\psi} \cm_j \psi$ by reading of the index $j$ in the tuple $\intchannelSet$ in \cref{eq:dirac_basis}. 
	As one can see from \cref{eq:part_bosonized}, non-vanishing $\langle \auxf_j\rangle$ give rise to dynamically generated mass terms for the fermion fields. 
	These mass terms spontaneously break different symmetries of the action \cref{eq:part_bosonized}. 
	The possibility of certain symmetry breaking schemes is, of course, dependent on the choice of interaction channels, i.e., on the set $\indSet$ (see the definition below \cref{eq:part_bosonized}).
	
	In \cref{app:symmetries}, we define the (chiral) symmetry group for $2+1$-dimensional fermion fields.
	Considering no interaction terms, the fermion fields contained in $\psi$ are invariant under transformations of the group $\U(4\N)$ composed of chiral transformation of the group $\hyperref[sym:U2N_ch]{\U_{\gamma}(2N)}$ and isospin transformations, which are elements of $\hyperref[sym:U2_iso]{\SU_{\vpauli}(2)}$.
	This invariance is not explicitly broken in the action \labelcref{eq:part_bosonized} (or \cref{eq:FFmodel}) if $\coupling_j = \lambda$ for\footnote{In fact, the full chiral symmetry group is already realized, when $\lambda_j = \lambda$ for $j=1,2,3$.} $j = 1, \ldots, 16$. 
	When a subgroup of the interaction channels is taken into account in the partially bosonized action \labelcref{eq:part_bosonized}, only subgroups of the chiral symmetry transformations might be realized (depending on the choice of $\indSet$). 
	The reminiscent symmetry transformations, which are relevant in \cref{sec:results}, are also defined in \cref{app:symmetries}.
	
	A mass term of the form $m \bar{\psi} \psi$, as spontaneously generated by a non-vanishing expectation value of $\phi_1=\sigma$, breaks the full symmetry group $\hyperref[sym:U2N_ch]{\U_{\gamma}(2N)}$ down to the $\hyperref[sym:U1_I]{\U_{\I_4}(N)} \times \hyperref[sym:U1_g45]{\U_{\gamma_{45}}(N)}$ subgroup of vector transformations.
	Mass terms of the form $\ii m_4 \bar{\psi} \gamma_4 \psi$ (generated by $\phi_2 = \eta_4$)  and $\ii m_5\bar{\psi} \gamma_5 \psi$ (generated by $\phi_3 = \eta_5$) have the same symmetry breaking pattern, but, in addition also break one of the two $\Z_2$ parity transformations\footnote{The parity transformation in $2+1$-dimensional Euclidean spacetime is defined as an inversion of an odd number of axes.
	Our convention is to flip all $3$ spacetime coordinates.
	The ambiguity of having two different parity transformations, that can act on the fermion fields, has its origin in the reducible spinor representation.}
	\begin{align}
		P_4: \quad\psi(\xstV) \rightarrow \gamma_4 \psi(-\xstV)\, &, \quad \bar{\psi}(\xstV) \rightarrow \bar{\psi}(-\xstV) \gamma_4; \label{sym:P4}\\
		P_5: \quad\psi(\xstV) \rightarrow \gamma_5 \psi(-\xstV)\, &, \quad \bar{\psi}(\xstV) \rightarrow \bar{\psi}(-\xstV) \gamma_5. \label{sym:P5}
	\end{align}
	In addition, there is a fourth mass term $m_{45} \bar{\psi} \gamma_{45} \psi$, which does not break a continuous symmetry, but both $P_4$ and $P_5$.
	This parity-odd mass can be dynamically generated by a non-vanishing expectation value of $\phi_4 = \eta_{45}$.
	These four different mass terms have an interpretation in condensed matter applications, such as graphene or similar systems \cite{PhysRevLett.53.2449, PhysRevLett.61.2015, PhysRevLett.98.186809}, see \Rcite{Hands:2021eyc} for a summary.
	We assign quantum numbers $+,-$ to the auxiliary bosonic fields in \cref{eq:phi_sym} according to their transformation behavior under the parity transformations \labelcref{sym:P4} and \labelcref{sym:P5}.
	The fields and their quantum numbers are listed in \cref{tab:P4P5_quantum_numbers}.
	Later, we will refer to the fields by their behavior under parity, e.g., $\eta_4$ and $\vec{\pi}_4$ are fields with quantum numbers $(-,+)$ and $\sigma, \vec{a}_0$ are $(+,+)$ fields. 
	\begin{table}[htbp]
		\begin{tabularx}{.5\columnwidth}{Y Y}
		\toprule
		$(P_4,P_5)$
		& $\auxf_j$	\\ 
		\hline
		$(+,+)$
		& $\sigma, \vec{a}_0$ \\
		\hline
		
		$(-,+)$
		& $\eta_4, \vec{\pi}_4$ \\
		\hline
		
		$(+,-)$
		& $\eta_5, \vec{\pi}_5$ \\
		\hline
		
		$(-,-)$
		& $\eta_{45}, \vec{\pi}_{45}$ \\
			
		\hhline{ = = } 	
		\end{tabularx}	
		\caption{\label{tab:P4P5_quantum_numbers} The quantum numbers $(P_4, P_5)$ of the fields $\auxf_j$ in \cref{eq:phi_sym}.}
	\end{table}
	
	After integration over the fermion fields in \cref{eq:part_bosonized}, one obtains an effective action, depending solely on the bosonic fields,
	\begin{align}
		\frac{\seff\left[\vauxf\right]}{\N} &= \int \dr^3x \, \sum_{j \in \indSet} \frac{\auxf_j^2 }{2 \coupling_j}  \ -  \  \Tr \ln \beta \D, \label{eq:Seff}
	\end{align}
	where $\D$ has been multiplied with $\beta$ in order to ensure a dimensionless argument of the logarithm. 
	This introduces only a temperature-dependent constant to the partition function.
	We recognize that $\seff$ is proportional to $N$.
	Taking the limit $\infN$ for the FF models is equivalent to a mean-field approximation at finite $\N$, which is what we consider for the rest of this work.
	In this approximation one takes the fermionic fluctuations fully into account through the functional $\Tr \ln$ over the Dirac operator $\D$, while  all quantum fluctuations in the bosonic degrees of freedom are neglected. 
	This causes the global minimum $\vauxf(\xsV) = \vminauxf(\xsV) $ of $\seff$ to be the only relevant contribution in the partition function.
	Thus, expectation values of observables can be computed by evaluating them on the respective global minimum.
	This can become problematic when the effective action has multiple, degenerate global minima, for example in the case of a first order phase-transition or minima which are related by symmetry transformations on the level of the effective action.
	In the latter case, one formally has to introduce a small symmetry breaking parameter $z$ and make the extrapolation $z \rightarrow 0$ in order to remove ambiguities.
	In the mean-field approximation, however, it is common (see, e.g., \Rcite{Asakawa:1989bq, Buballa:2003qv, Buballa:2014tba}) and more practical to pick one of the degenerate minima whenever facing this situation.
	In the case of a first order phase-transition or critical end point, we will refrain from evaluating any quantities that depend on the minimum of the effective action in order to avoid ambiguities.

\subsection{Yukawa models}\label{sec:I_Yukawa}
	In order to generalize our analysis of the \gls{ff} models, as defined in \cref{eq:FFmodel}, to corresponding Yukawa models in \cref{sec:stability}, we introduce their action in $2+1$-dimensional Euclidean spacetime as
	\begin{align}
	&\S_{Y}[\vYf] =  \frac{\seff[h\vYf]}{\N} \label{eq:Yukawa}
	+ \int  \mathrm{d}^3 x \left[ \frac{1}{2} \left(\partial_\nu \vYf(\xsV)\right) ^2 + \sum_{n > 1}\kappa_n \left( \sum_{j\in \indSet} \Yf_j^2(\xsV)\right)^{n}\right],  
	\end{align}
	where $\vYf = \left(\Yf_j\right)_{j\in \indSet}$ contains scalar fields of canonical dimension $\textrm{energy}^{1/2}$, $h$ is the Yukawa coupling, $\kappa_n$ are the couplings of the self-interaction terms between the fields $\vYf$ and $\seff$ is defined in \cref{eq:Seff}. 
	In the mean-field approximation, these models can be analyzed using the stability analysis in the same manner as the FF models, as is discussed in \cref{sec:Yukawa}. 
	Thus, quantum fluctuations of the fields $\vYf$ are neglected and observables are computed by evaluating them on the global minimum of $\S_{Y}$ using the same formalism as described above for the FF models. 
	A field $\Yf_j$, that by its interaction channel in $\seff$ corresponds to a field $\auxf_j$ in a \gls{ff} model, has the same parity quantum numbers defined in \cref{tab:P4P5_quantum_numbers} from $\auxf_j$.
	\cref{eq:Yukawa} defines a \gls{qm}-type of model. 
	One can think of the fermion fields contained in $\seff$ via the fermionic determinant as interacting through the exchange of the dynamical bosonic fields $\Yf_j$, which in the \gls{qm} model correspond to light mesons.
		
	We note that the Ward identity for the expectation values of the scalar fields $\Yf_j$, analogous to \cref{eq:Ward} for the auxiliary fields $\auxf_j$, contains additional contributions from the self-interaction and kinetic terms.
	Thus, the expectation value of $\Yf_j$ does not directly yield the expectation value of a fermion bilinear (although these can nevertheless be computed).
	However, the spatial homogeneity of the bosonic fields still directly leads to the absence of inhomogeneous condensates.
	Consequently, the investigation of \glspl{ip} in Yukawa models is possible by studying the stability of the fields $\Yf_j$ in the same way as we study the stability of $\auxf_j$ in the \gls{ff} models.  
	
\section{Stability analysis of models with a baryon chemical potential\label{sec:stability}}
	The stability analysis is a technique to determine whether a spatially homogeneous field configuration is unstable with respect to inhomogeneous perturbations.
	It follows from this instability that an inhomogeneous field configuration is energetically favored over the homogeneous expansion point.
	This technique has seen regular use in such investigations, e.g., in \Rcite{Wagner:2007he,Buballa:2020nsi, Buballa:2020xaa, Koenigstein:2021llr} and, thus, we recapitulate only the most relevant steps.
	We provide, however, a more detailed derivation in \cref{app:stab}. 
	Note that the central result of this work, i.e., showing the stability of homogeneous condensates against inhomogeneous perturbations in all models with Lorentz-(pseudo)scalar interaction channels in a general manner, is discussed below in \cref{sec:absence}.

	We start by considering homogeneous field configurations $\vHauxf$ and apply spatially inhomogeneous perturbations $\ptvauxf$ to them, i.e.,
	\begin{equation}
		\vauxf (\xsV) = \vHauxf  + \ptvauxf (\xsV),\label{eq:field_expansion}
	\end{equation}
	where the perturbations are of an arbitrary shape and assumed to be of an infinitesimal amplitude.
	Inserting this into the effective action \labelcref{eq:Seff} enables a systematic expansion of $\seff$ in powers of $\ptauxf$.
	The zeroth order contribution $\seff^{(0)}$ \labelcref{eq:seff0} is the so-called homogeneous effective potential and the leading order contribution $\seff^{(1)}$ \labelcref{eq:seff1} is proportional to the homogeneous gap equation (compare with \cref{eq:seff1Fourier,eq:gapequation}).
	Thus, $\seff^{(1)}$ vanishes if the homogeneous expansion point $\vHauxf$ corresponds to a solution of the gap equation.
	The second order contribution contains the Hessian of the effective action, which is found to be diagonal in momentum space, but not necessarily diagonal in field space.
	The diagonalization -- if possible -- is done via a change of the field basis $\vauxf \to \vdbf$.
	For the theory considered in \cref{eq:Seff}, where only a baryon chemical potential is present, the Hessian is already diagonal in field space, i.e., $\vdbf=\vauxf$, if we set all non-zero $\lambda_j$ to $\lambda$.
	Nevertheless, we use the ``new'' basis from here on in order to be consistent with the later analysis, where a proper diagonalization is indeed needed. 
	We then obtain as the second order contribution
	\begin{align}
		\frac{\seff^{(2)}}{N} ={}& \frac{\beta}{2}  \int \intMeasureOverPi{2}{q} \left[\sum_{j \in \indSet}\, |\ptftdbf_j(\sV{q})|^2\, \Gamma_{\dbf_j}^{(2)} \left(\Mass^2, \mu, T, q^2\right) \right]  \label{eq:Seff_2_diag}
	\end{align}
	where $\ptftdbf(\qsV)$ are the Fourier coefficients of the perturbations and the magnitude of the perturbation's momentum $q=|\qsV|$. The bosonic two-point function $\Gamma_{\dbf_j}^{(2)}\left(\Mass^2, \mu, T, q^2\right)$ is the curvature of the effective action with respect to $|\ptftdbf(\sV{q})|$.
	For the curvature we find the explicit form
	\begin{widetext}
	\begin{align}
		\Gamma_{\dbf_j}^{(2)} \left(\Mass^2, \mu, T, q^2\right) \label{eq:gamma2} 
		={}&\frac{1}{\coupling}
		+\frac{\Nbar}{\beta}\sum_n  \int \intMeasureOverPi{2}{p}\, \left(\frac{ -1 }{\tilde \nu_n^2 + \sV{p}^2+\Mass^2}  + \frac{1}{2} \frac{ q^2 + a_{\dbf_j} \,\Mass^2 }{[\tilde \nu_n^2 + \sV{p}^2+\Mass^2][\tilde \nu_n^2 + (\sV{p}+\sV{q})^2+\Mass^2]}\right)\equiv\\
		\equiv{}& \frac{1}{\lambda}   -\ell_1\left(\Mass^2, \mu, T\right) + \frac{1}{2}\left(q^2 + a_{\dbf_j} \,\Mass^2\right)\ell_2\left(\Mass^2, \mu, T, q^2\right) \equiv \nonumber \\ \equiv{}&  \frac{1}{\lambda}   -\ell_1\left(\Mass^2, \mu, T\right) + \EllTwoArgs{\dbf_j}{\Mass^2}{\mu}{T}{q^2}, \nonumber
	\end{align}
	where $\tilde \nu_n = (\nu_n-\ii\mu)$, the fermionic Matsubara frequencies $\nu_n = 2 \uppi (n - \tfrac{1}{2}) / \beta$  and
	\begin{align}
		\Mass^2\big(\vauxf\,\big)=\sum_{j \in \indSet }  \cmstar_j \cm_j(\auxf_j)^2\equiv \Mass^2 \quad \text{with} \quad 
		\cmstar_j= \begin{cases}
			c_j\ &\text{for } c_j=1, \vpauli\\
			-c_j \ &\text{otherwise } 
		\end{cases}. \label{eq:MAndCStar}
	\end{align}
	We find that $a_{\dbf_j} = a_+=4$ for fields $\dbf_j$ with parity quantum numbers $(P_4,P_5)=(+,+),(-,-)$ and $a_{\dbf_j} = a_-=0$ for fields with  $(P_4,P_5)=(+,-),(-,+)$.
	Accordingly, we find two possible momentum dependent contributions 
	\begin{align}
		L_{2, +}(\Mass^2, \mu, T, q^2) = \frac{1}{2} \left(q^2 +  4\Mass^2\right) \ell_2(\Mass^2, \mu, T, q^2)\ , \quad 
		L_{2, -} (\Mass^2, \mu, T, q^2) = \frac{1}{2} q^2  \ell_2(\Mass^2, \mu, T, q^2) \label{eq:L2pm}
	\end{align}
	to the two-point function.
	The integral $\ell_2$ at $T=0$ assumes the simple form
	\begin{align}
	 \ell_2(\Mass^2, \mu, T=0, q^2)=\frac{2}{\uppi q}
	\begin{cases} 
		0 \,,\quad &\mu^2 > \Mass^2 + q^2/4  \\
		\arctan \left(\frac{\sqrt{q^2 + 4(M^2-\mu^2)}}{2\mu}\right)\,,\ & \Mass^2\leq \mu^2 \leq \Mass^2 + q^2/4\   \\	
		\arctan \left(\frac{q}{2|\Mass|}\right) \,,\quad &\mu^2 <  \Mass^2
	\end{cases}	. \label{eq:ell2T0}
	\end{align}
	Further expressions for $\ell_1$ and $\ell_2$ for various cases of $\mu,T, M^2, q^2$ can be found in \cref{app:l1,app:l2}. 
	\end{widetext}

	\begin{figure*}
		\centering
		\includegraphics[width=0.95\linewidth]{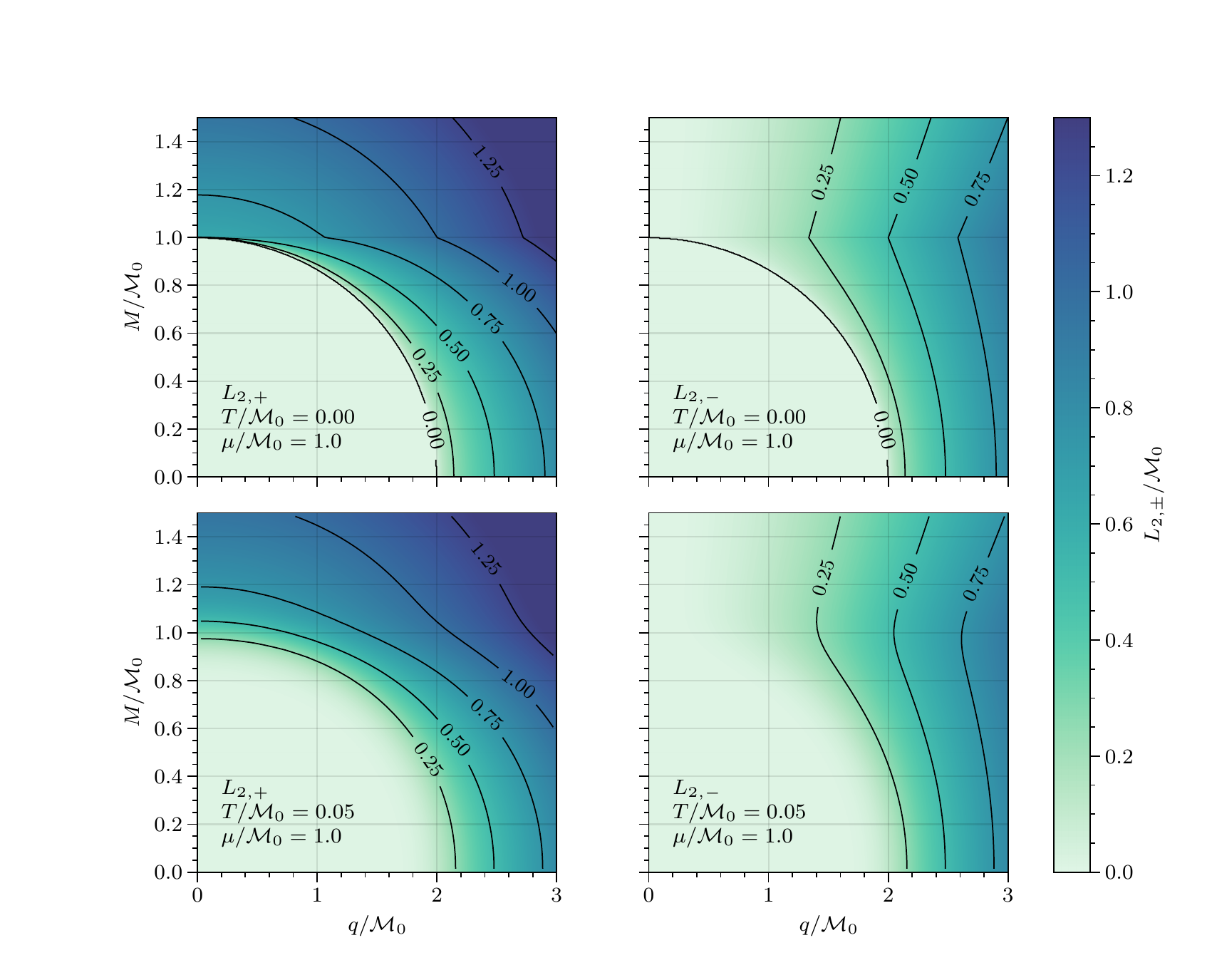}
		\caption{$\EllTwoPlus$ and $\EllTwoMinus$ for $T/\minMass_0=0,0.05$ and $\mu/\minMass_0=1.0$ as a function of $q$ and $\minMass$. Also at finite temperature $\EllTwoArgs{-}{\Mass^2}{\mu}{T}{q^2=0}=0$. However, the $0.0$ contour line and its label are not drawn, because they would be obscured by the axis. In the ancillary files to this paper, we provide a script, which allows to do the numerical computation of $\EllTwoPlusMinus$ for arbitrary values of $(\Mass^2, \mu, T, q^2)$ and that produces this figure. See \cref{eq:gamma2,eq:L2pm} for the definition of $\EllTwoPlusMinus$.}
		\label{fig:ell2}
	\end{figure*}
	
	In order to make conclusions about whether inhomogeneous condensates are favored over homogeneous ones, we use the global homogeneous minimum as the homogeneous expansion point, i.e., the field configuration $\vminHauxf$ that minimizes $\seff^{(0)}$ globally with the corresponding mass $\minMass^2 := \Mass^2(\vminHauxf)$. 
	This field configuration is also a solution of the gap equation \labelcref{eq:gapequation}. 
	Therefore, $\seff^{(0)}$ assumes its minimal value while $\seff^{(1)}$ vanishes (since it is proportional to the gap equation, compare with \cref{eq:seff1Fourier,eq:gapequation}). 
	Thus, negative values of $\Gamma_{\dbf_j}^{(2)}(\minMass^2, \mu, T, q^2)$ will signal that there is an even deeper minimum in the direction of $\ptftdbf_j(\sV{q})$.
	In order to provide dimensionless quantities, we express all parameters in units of the mass $\minMass_0$, which is the mass corresponding to the global homogeneous minimum in the vacuum of the theory.

	\subsection{Absence of instability  \label{sec:absence}}	
	The global homogeneous minimum, that is the only relevant expansion point when searching for an \gls{ip}, is stable against homogeneous perturbations, i.e., $\Gamma_{\dbf_j}^{(2)}(\minMass^2, \mu, T, q^2=0) \geq 0 \ \forall \dbf_j$.
	Consequently, negative values for finite $q^2$ and, thus, an instability against inhomogeneous perturbations can be ruled out, if the $\sV{q}$-dependent part $L_{2, \dbf_j}$ of $\Gamma_{\dbf_j}^{(2)}(\minMass^2, \mu, T, q^2)$ is a monotonically increasing function of $q^2$.
	This is exactly the case for $\EllTwoPlusMinus$.
	For $T=0$ the analytical form of $\EllTwoPlusMinus$ given by \cref{eq:L2pm,eq:ell2T0} reveals its monotonically increasing behavior for all $\minMass^2,\mu,q^2$.
	The same is true for finite temperatures which can be verified by numerical calculations of $\ell_2$.
	In \cref{fig:ell2}, the functional behavior of $\EllTwoPlusMinus$ for $\mu/\minMass_0=1.0$ and $T/\minMass_0=0.0, 0.05$ is plotted in a color plot.
	We conclude that the general model defined by \cref{eq:Seff} does not experience an instability towards an \gls{ip}.  
	This conclusion holds when considering any set of interaction channels as given by $\indSet$.
	Moreover, these models also do not feature a so-called moat regime \cite{Pisarski:2021qof}. 
	This regime is defined by a negative wave-function renormalization $Z \propto \dr^2 \Gamma^{(2)} / \dr q^2$ and is often an accompanying phenomenon to an \gls{ip} but can also exist independently \cite{Koenigstein:2021llr}.
	Our computations show that $\EllTwoPlusMinus$ always yields $Z \geq 0$, since they are monotonically increasing functions in $q$.  
	
	Additionally, we note that $\ell_2 \sim \Theta\left(\mu^2 - \Mass^2 - q^2/4\right)$ at $T=0$, which causes the two-point function to be constant for momenta $0 < q^2/4 < \mu^2 - \Mass^2$. 
	Here not only the wave function renormalization $Z$ vanishes, but also any higher derivative of $\Gamma^{(2)}$ with respect to $q$.
	A special point is at $\mu/\minMass_0=1$ and $T=0$, where the homogeneous phase transition in the $2+1$-dimensional \gls{gn} is often regarded as being first order.
	However, the system rather exhibits a critical endpoint at this point \cite{Inagaki:1994ec}. 
	This means that the effective potential (e.g.\ computed in \Rcite{Klimenko:1987gi, Buballa:2020nsi}) is flat between some homogeneous field values with $\Mass \in [0.0, 1.0]$.
	Such a flatness is also observed in the two-point function where the contribution $1/\lambda - \ell_1$ vanishes, which causes the two-point function to be constant zero for momenta $0 < q^2/4 < \mu^2 - \Mass^2$. 
	This vanishing curvature of the effective action is a hint for a degeneracy between homogeneous and inhomogeneous condensates.
	Such a behavior has already been observed in the case of the $2+1$-dimensional \gls{gn} model also at $\mu/\minMass_0 = 1.0, T/\minMass_0 = 0$, which has been revealed by a study \cite{Urlichs:2007zz} with an one-dimensional ansatz for the chiral condensate. 
	We expect this degeneracy to be restricted to $(\mu, T)/\minMass_0 = (1.0,0.0)$ and, especially, the homogeneous condensates to be favored against inhomogeneous ones for all non-vanishing $T$.
	The analysis in \Rcite{Urlichs:2007zz} suggests that also higher orders of the expansion of the effective action would vanish for a certain range of finite momenta. 
	In \cref{sec:res_FF}, we show that the homogeneous phase diagram of all possible models described by \cref{eq:Seff} is identical to the one of the $2+1$-dimensional \gls{gn} model.
	Thus, the above discussion of the flatness of the effective potential and the two-point function also applies to these models.
	
	It should be noted that the absence of an instability does not completely rule out the existence of an \gls{ip}.
	As discussed in \Rcite{Koenigstein:2021llr} at the example of the $1+1$-dimensional \gls{gn} model, it is possible for the homogeneous minimum and global inhomogeneous minimum to be separated by a energy barrier.
	Here, the homogeneous minimum appears stable against inhomogeneous perturbations even though an \gls{ip} is energetically favored.
	Such a phase can only be found by calculations with a guess of ansatz functions for the condensates or by explicit numerical brute-force minimizations using lattice field theory.
	For the $2+1$-dimensional \gls{gn} model, there is evidence in the literature that this is not realized.
	Lattice minimizations of this model have not found any other \gls{ip} at finite lattice spacings than the ones also obtained by a stability analysis of the lattice regularized models \cite{Buballa:2020nsi, Pannullo:2021edr, Winstel:2022jkk}. 
	It is important to note that these \glspl{ip} vanish when taking the continuum limit \cite{Narayanan:2020uqt, Buballa:2020nsi}, as the bosonic two-point function converges towards a momentum dependence proportional to $\EllTwoPlus$.
	These results of the $2+1$-dimensional \gls{gn} model suggest that the stability analysis applied to our general \gls{ff} model \labelcref{eq:Seff} also does not miss an \gls{ip}.
	Also, to our knowledge all \glspl{ip}, which are observed in model investigations, can at least in some parameter region be detected by the stability analysis.
	In these studies, one finds a second order phase transition between the \gls{ip} and the chirally symmetric phase. 
	At least, this second-order transition can always be detected by analyzing the stability of the symmetric minimum of the effective potential.
	The sum of these arguments combined with our analysis is strong evidence for the absence of inhomogeneous condensates in all of the models described by \cref{eq:Seff} (or, equivalently, \cref{eq:part_bosonized}) independent of the considered (sub)set of interaction channels given by $\indSet$. \\
	
	Since this result is obtained in the mean-field approximation, one needs to consider its predictive power for the full quantum theories. It is typically found that bosonic fluctuations tend to disfavor and/or disorder (in-)homogeneous condensation\footnote{ \Rcite{landauStatisticalPhysicsPart1980} provides a rather general argumentation which excludes the spontaneous breaking of a discrete symmetry in one dimension, which is backed by the results in \Rcite{Stoll:2021ori} that found no homogeneous condensation in this model at $T\neq 0$. The breaking of a continuous symmetry such as the translational symmetry is forbidden in $1+1$ dimensions at finite temperature by the no-go theorem presented in \Rcite{Mermin:1966fe}. Thus, we expect that also inhomogeneous phases are disordered by the bosonic fluctuations. In \Rcite{Lenz:2020bxk}, the authors cannot distinguish between a long-range order scenario and an \gls{ip} in the $(1+1)$-dimensional \gls{gn} model.} \cite{Scherer:2012fjq, Lenz:2020bxk, Lenz:2021kzo, Stoll:2021ori,Lenz:2023wvk, Ciccone:2022zkg}.
	Based on these findings, we expect that the non-existence of \glspl{ip} in $2+1$-dimensional \gls{ff} and Yukawa models in the mean-field approximation is a clear signal that inhomogeneous ground states are not present in the corresponding full quantum theories.
	A scenario with an inhomogeneous ground state in a full quantum theory which is not present in the mean-field approximation has -- to our knowledge -- never been observed.		
	In $2+1$-dimensional \gls{ff} models, one has so far only seen oscillating but also damped correlation functions for the auxiliary fields \cite{Strouthos:2003js}. 
	However, these lattice results \cite{Strouthos:2003js} have been obtained on rather crude lattices. So, we expect these oscillations to be reminiscent of the \gls{ip}, which can be found in mean-field investigations of \gls{ff} models at finite lattice spacings, but vanish in the continuum limit \cite{Buballa:2020nsi, Narayanan:2020uqt}.

\subsection{Generalization to Yukawa models\label{sec:Yukawa}}
	When the models are only subjected a baryon chemical potential, it is rather straightforward to generalize the stability analysis of \gls{ff} models to Yukawa models, which are defined as in \cref{eq:Yukawa}.
	We only discuss the final result here and a more detailed derivation can be found in \cref{App:YukawaStab}.
	
	Applying the stability analysis to the effective action as defined in \cref{eq:Yukawa}, the second order contribution again is the first non-vanishing correction to the homogeneous action $S_{Y}^{(0)}$, when the homogeneous expansion point is a solution of the gap equation.
	We find the second order contribution to the effective action
	\begin{align}
		\frac{\S_{Y}^{(2)}}{N} ={}& \frac{\beta}{2}  \int \frac{\mathrm{d}^2 q}{(2\uppi)^2}\, \left[\sum_{j \in \indSet}\, |\ptftdbYf_j(\mathbf{q})|^2\, \Gamma_{\dbYf_j}^{(2)} (q^2) \right] \label{eq:seff2Yukawa}, 
	\end{align}	
	where $\dbYf$ is the field basis that was used in the diagonalization of the Hessian\footnote{In contrast to the ordinary \gls{ff} models, it is possible for the Yukawa models with a baryon chemical potential to exhibit off-diagonal terms in the Hessian matrix.
		These contributions are, however, independent of $\sV{q}$ and can be removed by using symmetry transformations on the homogeneous expansion point.}. The two-point function
	\begin{align}
		\Gamma_{\dbYf_j}^{(2)}  \left(\Mass^2, \mu, T, q^2\right) ={}& \frac{1}{\lambda} -\ell_1\left(\Mass^2, \mu, T\right) +\label{eq:gtwoYukawaDiag}  \EllTwoArgs{\dbYf_j}{\Mass^2}{\mu}{T}{q^2} +  \\
		&+ \frac{1}{2} q^2 + \sum_{n >1 }\kappa_n n \left[ 2 \left(\vHdbYf^2\right)^{n-1} + 4(n-1)  \HdbYf_j^2\left(\vHdbYf^2\right)^{n-2} \right] \nonumber
	\end{align}
	can now be identified as the curvature of the effective action in the field direction $\ptftdbYf_j(\qsV)$. 
	Again, one finds either $\EllTwoPlus$ or $\EllTwoMinus$ as the momentum dependence of the two-point function.
	As is obvious from \cref{eq:gtwoYukawaDiag}, the additional contributions compared to the \gls{ff} two-point function are either constant or monotonically increasing in $q^2$.
	Thus, by the reasoning given in the last section, these additional terms cannot facilitate the appearance of an instability towards an \gls{ip} since the corresponding \gls{ff} model does not exhibit such an instability for any $\vminHauxf,\mu,T,q^2$ already.

\section{Results of the analysis for specific models}
\label{sec:results}
	In this section, we present examples of \gls{ff} and Yukawa models where the condensates do not develop an instability towards inhomogeneous perturbations and are, thus, very unlikely to feature an \gls{ip}.
	In \cref{sec:res_one_chempot}, \gls{ff} models with only a baryon chemical potential but multiple interaction channels (see \cref{eq:FFmodel}) are presented.
	These results are obtained using the stability analysis, as explained in detail in \cref{sec:stability}, where also the reasoning for the absence of instabilities is explained in detail. 
	After that, we allow for multiple chemical potentials in \cref{sec:res_mult_chempot} and explain the differences of the analysis compared to only a baryon chemical potential. Again, examples for model calculations are discussed.
	Finally, we turn towards the extension of our findings to Yukawa models in \cref{sec:res_Yukawa}.
	\subsection{Four-fermion models \label{sec:res_FF}} 
	Before we turn towards the stability analysis of the bosonic two-point functions, we shortly present our finding for the homogeneous phase diagram of \gls{ff} models described by \cref{eq:FFmodel,eq:part_bosonized} (by considering different interactions channels in the set $\indSet$).
	 
	The homogeneous phase diagram of the $2+1$-dimensional \gls{gn} model, i.e., the phase diagram when restricting the field to homogeneous field configurations, as first obtained in \Rcite{Klimenko:1987gi}, features a phase at low temperature and chemical potential where the discrete chiral symmetry is spontaneously broken by a non-zero chiral condensate \cite{Klimenko:1987gi, Rosenstein:1988pt}.
	At finite temperature this phase is separated from the chirally symmetric phase by a second order phase transition. At $(\mu,T)/\minMass_0=(1.0,0.0)$ one obtains a first order transition point, where degenerate minima with $\minMass/\minMass_0 \in [0.0, 1.0]$ have the same minimal effective potential, which becomes flat.
		
	The homogeneous phase diagram for all possible models, that can be obtained from \cref{eq:FFmodel} by setting certain couplings either to zero or $\coupling$, is identical to that of the \gls{gn} model.
	This result is related to the present symmetries of the action, which allow to pick homogeneous minima, where only the scalar channel $\sigma$ develops a non-zero expectation value (note that this is only possible when the fields are restricted to being homogeneous).
	Other homogeneous, global minima are connected to this one via a (chiral) symmetry transformation. 
	Exempt from this are the fields with parity quantum numbers $(-,-)$ namely $\bar\eta_{45}$ and $\vec{\pi}_{45}$. 
	By analyzing the homogeneous effective potential, we can nevertheless show that the homogeneous fields $\bar\eta_{45}, \vec{\pi}_{45}$ develop vanishing expectation values for all temperatures and baryon chemical potentials.
	First, we discuss this allowing for all isoscalar channels, which yields the introduction of the auxiliary fields $\sigma, \eta_4, \eta_5,\eta_{45}$.
	Using the symmetries of \cref{eq:Seff} (or equivalently \cref{eq:part_bosonized}) one can use the chiral transformations \cref{sym:U2N_ch} to obtain non-vanishing expectation values only for $\sigma$ and $\eta_{45}$,  
	Then, we analyze the fermion contribution $\Tr \ln \Q$, which can be represented in a block-diagonal form with $2\times 2$ blocks corresponding to the irreducible spin representation in $2+1$ dimensions. 
	These blocks can be interpreted as \gls{gn} model contributions in the irreducible fermion representation with chemical potential $\mu$ and homogeneous fields $\Hom{\phi}_{L/R} \equiv \Hom{\sigma} \pm \Hom{\eta}_{45}$ (see the theory sections of \Rcite{Buballa:2020nsi, Pannullo:2021edr} for a more in-depth discussion of these irreducible blocks).
	The model decomposes into two \gls{gn} models with the same chemical potentials, which gives us $\Hom{\phi}_L = \Hom{\phi}_R$ and, thus $\Hom{\eta}_{45} = 0$.
	An analogous analysis involving some additional isospin rotations is valid for $\vec{\pi}_{45}$. 
	Thus, we do not observe spontaneous parity breaking in the \gls{ff} models for all $\mu$ and $T$.
		
	The order of the homogeneous phase transition might be relevant, since the stability analysis of the bosonic two-point function in the $1+1$-dimensional \gls{gn} revealed that a first order phase transition can cause stability of the homogeneous minimum even though there is an energetically preferred \gls{ip} \cite{Koenigstein:2021llr}.
	Here, the global energetically preferred inhomogeneous minimum was only connected to the symmetric homogeneous field configuration which was not energetically favored over the stable non-zero minimum.
	The fact that in $2+1$ dimensions the critical point occurs at a single point at zero temperature increases the confidence in the presented results obtained via the stability analysis.
	Moreover, by inspecting all possible expansion points via varying the mass $\Mass$, we can rule out a scenario as in $1+1$ dimensions, since in $2+1$ dimensions there is no other unstable expansion point while the physically relevant expansion point $\minMass$ is stable.

\subsubsection{Models with baryon chemical potential\label{sec:res_one_chempot}}
			
	\cref{tab:Results} summarizes the central findings from the stability analysis for some models that are obtained from \cref{eq:part_bosonized} by defining $\indSet$ and setting $\lambda_j = \lambda, \, j \in \indSet$.
	The specific models are chosen with respect to their relevance to phenomenology and the literature.
	The two-point functions for models with other combinations of channels can easily be calculated as discussed in \cref{sec:stability}. 
	The respective symmetry transformations of the effective actions allow to obtain homogeneous expansion points $\Hauxf_j$, which are vanishing except for $\Hom{\sigma}$. 
	None of the two-point functions in these models exhibit a different momentum dependence than $\EllTwoPlusMinus$.
	As discussed in \cref{sec:stability} and illustrated by \cref{fig:ell2}, $\EllTwoPlusMinus$ are positive, monotonically increasing functions of the square of the momentum $\qsV$ of the inhomogeneous perturbations. 
	Thus, we expect the absence of \glspl{ip} and moat regimes in all of these models, as discussed in detail in \cref{sec:absence}. 
	Therein, we also explain that there is the possibility of a degeneracy between the homogeneous condensate and the inhomogeneous condensate, as observed at $(\mu, T)/\minMass_0 = (1.0,0.0)$ in the \gls{gn} model \cite{Urlichs:2007zz}.
	By our analysis, we expect this degeneracy to be restricted to $(\mu, T)/\minMass_0 = (1.0,0.0)$ and, especially, the homogeneous condensates to be favored against inhomogeneous ones for all non-vanishing $T$.

	In \cref{tab:Results}, we discuss four \gls{ff} models explicitly and give their respective symmetry groups (defined in \cref{app:symmetries}) and indicate whether the momentum dependence of the two-point functions are $\EllTwoPlus$ or $\EllTwoMinus$. 
	
	The first row shows the renowned \gls{gn} model with one single isoscalar channel with parity quantum number $(+,+)$ (see \cref{sym:P4} and \cref{sym:P5} for the definition of the parity in $2+1$ dimensions and \cref{tab:P4P5_quantum_numbers} for the quantum numbers of the fields).
	The two-point function exhibits the $\EllTwoPlus$ dependence, which was already documented in \Rcite{Buballa:2020nsi}.
	
	The second row shows the $2+1$-dimensional analog of the $3+1$-dimensional \gls{njl} model. 
	It breaks the axial symmetry transformations \cref{sym:U1_g4,sym:U1_g5}, leaving only the combined isospin and axial symmetries \cref{sym:SU2_A_g4,sym:SU2_A_g5} as chiral transformations.
	Due to the ambiguity of the $\gamma_{5}$ operator in $2+1$ dimensions, it makes sense to use the generators of both transformations to construct an ``\gls{njl}" Lagrangian.
	Similar to the case in $3+1$ dimensions, we find that the momentum dependence of the two-point functions of the  $\pi$-fields are given by $\EllTwoMinus$, while it is $\EllTwoPlus$ for the $\sigma$ field \cite{Buballa:2020xaa, Pannullo:2022eqh}. 
	
	The third row shows the $2+1$-dimensional \gls{chign} model \cite{Gracey:2018cff} with an additional $\left(\bar{\psi} \gamma_{45} \psi\right)^2$ interaction term that is special to $2+1$ dimensions\footnote{The analysis of this model with respect to an \gls{ip} and their disordering at finite flavor numbers is also motivated in \Rcite{Pisarski:2020dnx}.}. 
	Again, due to there being two axial transformations \cref{sym:U1_g4,sym:U1_g5}, a Lagrangian involving both generators is an appropriate choice. 
	A non-zero bosonic field $\eta_{45}$ breaks both parity symmetries \cref{sym:P4,sym:P5} spontaneously, but does not break one of the continuous chiral symmetries.
	Due to $\bar\eta_{45}=0$ for all temperatures and baryon chemical potentials, the two-point functions of the fields do not mix (as the only off-diagonal terms are $\propto \bar\eta_{45}$, compare with \cref{eq:l2_tr}) and the diagonalizing field basis coincides with the auxiliary bosonic fields introduced in the bosonization.
	Thus, one of the two parity transformations could only be spontaneously broken if one of the fields with a negative parity quantum number (compare \cref{tab:P4P5_quantum_numbers}) develops a spatial modulation. 
	However, also for this model we found that the momentum dependent part of the two-point functions of all present bosonic fields is either $\EllTwoPlus$ or $\EllTwoMinus$ and, thus, homogeneous condensates never develop an instability towards a spatially dependent condensate.
	
	The last row lists what we call the \gls{cnjl} model, since it features all interaction channels present in \cref{sec:theory,eq:FFmodel}.
	Again, note that $\bar\eta_{45}=\vec{ \bar\pi}_{45}=0$ which prevents off-diagonal second order terms from the fermionic contribution, see \cref{eq:seff2_nondiag}.
	Even though we considered by far the largest amount of interaction channels in this model, no two-point function with a different momentum dependence than $\EllTwoPlusMinus$ is obtained.
		
	As discussed in \cref{sec:absence}, two-point functions for fields, which have mixed parity quantum numbers ($(P_4, P_5) = (+,-)$ or $(P_4, P_5) = (-,+)$ ), have a momentum dependence proportional to $\EllTwoMinus$, while the others have a momentum dependence proportional to $\EllTwoPlus$.
	This is independent of how large the symmetry group is and of the number of interaction channels considered. 
	Even when considering all $16$ Lorentz-scalar \gls{ff} interactions (see \cref{eq:dirac_basis}), we do not see a different mathematical structure of the two-point functions. 
	
	Therefore, according to the argument given in \cref{sec:absence}, none of the models, which can be described by \cref{eq:part_bosonized} by defining $\indSet$ and setting $\lambda_j = \lambda, \, j \in \indSet$, exhibits an instability towards an \gls{ip} for any $\minMass^2, \mu, T, \sV{q}^2$. 
	As argued in detail in \cref{sec:absence}, this is strong evidence for the absence of \glspl{ip} in these models.
		
	\begin{table*}[tb]
		\begin{tabularx}{\textwidth}{Y Y Y Y Y Y  }
			\toprule
			Model
			& Used channels $\cm_j$ 
			& Field basis $\vdbf_j$ diagonalizing $\seff^{(2)}$
			& \multicolumn{2}{c}{\parbox{2.5cm}{Momentum dependence of $\gtwo_{\varphi_j}$} } 
			& Symmetry groups\\
			\cline{4-5}
			&
			&
			& $\EllTwoPlus$
			& $\EllTwoMinus$
			&\\
			\cline{1-6}
			\gls{gn}
			& $1$
			& $\sigma$ 
			&$\sigma$
			& 
			& $\hyperref[sym:U1_I]{\U_{\I_4}(N)} \times \hyperref[sym:U1_g45]{\U_{\gamma_{45}}(N)} \times \hyperref[sym:Z2_g5]{\Z_{\gamma_{5}}(2)} \times \hyperref[sym:U2_iso]{\SU_{\vpauli}(2)}\times \hyperref[sym:P4]{P_4} \times \hyperref[sym:P5]{P_5}$\\ 
			\cline{1-6}
			\gls{njl}
			&$1, \ii \vpauli \gamma_4, \ii \vpauli\gamma_5$                                
			& $\sigma, \ve{\pi}_4, \ve{\pi}_5$  
			& $\sigma$  
			& $\ve{\pi}_4, \ve{\pi}_5$
			& $\hyperref[sym:U1_I]{\U_{\I_4}(N)} \times \hyperref[sym:U1_g45]{\U_{\gamma_{45}}(N)}  \times \hyperref[sym:SU2_A_g4]{\SU_{A, \gamma_{4}}(2N)}\times \hyperref[sym:SU2_A_g5]{\SU_{A, \gamma_{5}}(2N)} \times \hyperref[sym:U2_iso]{\SU_{\vpauli}(2)}\times \hyperref[sym:P4]{P_4} \times \hyperref[sym:P5]{P_5}$  \\
			\cline{1-6}
			\gls{chign}$_P$
			& $1, \ii \gamma_4, \ii \gamma_5, \gamma_{45}$                                 
			& $\sigma, \eta_4, \eta_5, \eta_{45}$ (for $\bar\eta_{45} = 0$)  
			& $\sigma,\eta_{45}$
			& $\eta_4,\eta_{5}$
			& $\hyperref[sym:U2N_ch]{\U_{\gamma}(2N)} \times \hyperref[sym:U2_iso]{\SU_{\vpauli}(2)} \times \hyperref[sym:P4]{P_4} \times \hyperref[sym:P5]{P_5}$ \\
			\cline{1-6}
			\gls{cnjl}
			&$1, \ii \gamma_4, \ii \gamma_5, \gamma_{45},$  $ \vpauli, \ii \vpauli \gamma_4, \ii \vpauli\gamma_5, \ii \vpauli \gamma_{45}$                             
			&  $\sigma, \eta_4, \eta_5, \eta_{45}$                    $\ve{a}_0, \ve{\pi}_4, \ve{\pi}_5, \ve{\pi}_{45}$   ( for $\bar\eta_{45} = \bar{\ve{\pi}}_{45}= 0$ ) 
			& $\sigma,  {\eta_{45}},{\ve{\varsigma}},{\ve{\pi}_{45}}$ 
			&${\eta_4},  {\eta_{5}}, {\ve{\pi}_4},  {\ve{\pi}_{5}}$
			& $\hyperref[sym:U2N_ch]{\U_{\gamma}(2N)} \times \hyperref[sym:U2_iso]{\SU_{\vpauli}(2)} \times \hyperref[sym:P4]{P_4} \times \hyperref[sym:P5]{P_5}$ \\
			\botrule
		\end{tabularx}
		\caption{Stability analysis of the bosonized \gls{ff} models. We allow for finite baryon chemical potential $\mu$ and finite temperature $T$. The first column gives the models abbreviations for their names for further reference (whenever available names existing in the literature are used). In the second column, the respective interaction channels kept from \cref{eq:FFmodel} are listed. The rest is removed by setting $\coupling_k = 0$. The third column lists the field basis $\dbf_j$, for which the \cref{eq:seff2} can be diagonalized and, thus, a meaningful stability analysis can be performed. The fourth column indicates whether the momentum dependence of $\gtwo_{\dbf_j}$ is given by $\EllTwoPlus(\Mass^2, \mu, T, \qsV^2)$ or $\EllTwoMinus(\Mass^2, \mu, T, \qsV^2)$. The fifth column gives the full symmetry group of the model. The groups are clickable and refer to the definition of the symmetry group.}
		\label{tab:Results}
	\end{table*}

\subsubsection{Models with multiple chemical potentials \label{sec:res_mult_chempot}}
	In this section, we allow for multiple chemical potentials in \gls{ff} models in addition to the baryon chemical potential, which is introduced in \cref{eq:FFmodel} via the usual $\mu \bar{\psi} \gamma_3\psi$ term. 
	Namely, we will study the effect of finite isospin chemical potential, introduced with the term $\mu_I \bar{\psi} \gamma_3 \tau_3\psi$, and chiral chemical potential, that is introduced as $\mu_{45} \bar{\psi} \gamma_3 \gamma_{45} \psi $.
	Although this requires an extensive analysis, we can again identify the well-known function $\EllTwoPlus$ in the two-point functions. 
	Thus, also the models discussed below do not develop an instability towards an \gls{ip}.
	Also, the existence of a moat regime with a negative wave-function renormalization can be ruled out, as $Z \geq 0$ according to the discussion in \cref{sec:stability}.
	
	In general, the introduction of multiple chemical potentials induce that (homogeneous) expectation values of several of the auxiliary fields $\Hauxf_i$ will have non-zero expectation values, e.g., studying finite $\mu_{45}$ can lead to a non-vanishing expectation value of $\Hom{\eta}_{45}$.
	Also, the homogeneous phase diagram has to be computed separately for each individual case and can have a more involved phase structure (compare, e.g., \Rcite{Pannullo:2021edr}).
	
	This also significantly complicates the analysis of the bosonic two-point functions, although in principle the method can be applied as introduced in \cref{sec:stability}. 
	The main difficulty is the diagonalization of the bracketed field space matrix in \cref{eq:seff2_nondiag}, which becomes quite involved depending on the studied action.
	For a large number of interaction channels and multiple non-vanishing $\Hauxf_i$ it can become even impossible to diagonalize this expression. 
	Thus, a general analysis for multiple theories, as in \cref{sec:stability} for \gls{ff} models with baryon chemical potential, cannot be presented by us.
	
	However, there are still some combinations of interaction channels and chemical potentials, where one can still obtain the momentum dependence of the two-point functions by a comparatively easy diagonalization obtained by a suitable choice of field coordinates $\dbf_j$.   
	For these models, we find the fields $\dbf_j$ such that the fermion propagator \labelcref{eq:propagator_hom} is block-diagonal with $2\times2$ blocks, where each corresponds to an irreducible spinor representation.
	Then, the contribution of the fermion-loop (as written down in, e.g., \cref{eq:trace_ferm_loop,eq:one_loop_FF_secorder}) also decomposes when $\ptD$ is written as a function of the $\ptdbf_j$.
	The vertices\footnote{By vertices, we mean the $8\times8$ matrices, that describe the coupling between the bosonic field to the spin/isospin degrees of freedom of the fermion fields.} corresponding to the interaction of the fermions with the new field basis $\dbf_j$ project out either one or multiple of the $2\times2$ blocks and diagonalize the bracketed field space matrix in \cref{eq:seff2_nondiag}.
	Then one can proceed with the analysis as described in \cref{sec:stability,app:stab}.
	
	\cref{tab:Results_chemipotentials} summarizes the models, where such an analysis with respect to the stability of homogeneous condensates against inhomogeneous perturbations was performed by us.
	All two-point functions obtained in the models in \cref{tab:Results_chemipotentials} are proportional to $\EllTwoPlus$. 
	This originates in the circumstance, that in our derivation of the two-point function the field variables $\dbf_j$ are chosen such that one obtains a block-diagonal structure of the homogeneous fermion propagator \labelcref{eq:propagator_hom}.
	These blocks behave as $2+1$-dimensional \gls{gn} models (see \cref{tab:Results} for the momentum dependence of the two-point function of the $\sigma$ field in the \gls{gn} model) with a single effective chemical potential and a single field $\dbf_j$ or sums of the fields. 
	The effective chemical potentials are linear combinations of $\mu, \mu_{45}, \mu_{I}$.
	The momentum-dependence of the obtained two-point functions is given by linear combinations of $\EllTwoPlus$ with $\Mass^2$ and effective chemical potentials. 
	
	The model in the first row is a \gls{gn} model with an additional $\left(\bar{\psi} \gamma_{45} \psi\right)^2$ interaction channel and a chiral imbalance via finite $\mu_{45}$.
	A standard \gls{gn} model subjected to $\mu_{45}$ without the additional interaction channel has been studied in \Rcite{Pannullo:2021edr}. 
	The corresponding homogeneous expectation value of the field $\eta_{45}$, as induced by finite $\mu_{45}$ breaks parity spontaneously.
	However, the new field basis $\phi_{L/R}$ effectively decouples the theory into two, independent \gls{gn} models with chemical potentials $\mu \pm \mu_{45}$.
	This leads to two ``\gls{gn}-like'' two-point functions each with one independent field and chemical potential.
	The same procedure can be performed for the model in the second row, where instead of a $\left(\bar{\psi} \gamma_{45} \psi\right)^2$ the isospin channel $\left(\bar{\psi} \tau_3 \psi\right)^2$ (leading to an auxiliary bosonic field $a_{0,3}$ in the bosonic theory) is added to the \gls{gn} interaction and an isospin chemical potential $\mu_{I}$ introduces an isospin imbalance.
	
	In the third row, we allow for both isospin and chiral imbalance introducing both corresponding chemical potentials in addition to the baryon chemical potential $\mu$. 
	A $\left(\bar{\psi} \gamma_{45} \tau_3 \psi\right)^2$ interaction channel is studied in addition to the $\left(\bar{\psi} \psi\right)^2 $ interaction. 
	The corresponding bosonic auxiliary fields $\pi_{45, 3}$ and $\sigma$ can again be combined via linear combination to obtain the diagonalizing field basis $\varphi_{\pm}$. 
	The two-point functions $\gtwo_{\varphi_{\pm}}$ are now sums of different $\EllTwoPlus$ contributions, each with $\Mass^2 = \varPhi_{\pm}^2$, but with differing linear combinations of the chemical potentials dictating the exact form of $\EllTwoPlus$.
	Again, we refer to \Rcite{Pannullo:2021edr} where an analogous competition between two chemical potentials is studied in the phase diagram and, although this is not explicitly computed therein, in the two-point function of the $\sigma$ field.
	
	In the forth row, the most involved model in \cref{tab:Results_chemipotentials} is considered containing both previously introduced interactions resulting in the presence of the auxiliary bosonic field $\eta_{45}$ and $a_{0,3}$ in addition to the $\sigma$ channel simultaneously at a finite baryon, isospin and chiral density.
	We find the diagonalizing field basis to be $\phi_L, \phi_R, a_{0,3}$. 
	The momentum dependence of the two-point functions of $\phi_{L/R}$ are again sums of two $\EllTwoPlus$ contributions each with different linear combinations of the chemical potentials and $\Mass^2 = \left(\Hom{\phi}_{L/R} \pm \Hom{a}_{0,3}\right)^2$.
	However, $\gtwo_{a_{0,3}}$ is proportional to the sum of $\gtwo_{\phi_L}$ and $\gtwo_{\phi_R}$, combining four different contributions each proportional to $\EllTwoPlus$.
	
	Concluding, all the studied \gls{ff} models in \cref{tab:Results_chemipotentials} do not exhibit an instability of the homogeneous ground state when subjected to inhomogeneous perturbations and, thus, it is very unlikely that they feature an \gls{ip} (c.f.~\Rcite{Buballa:2020nsi, Narayanan:2020uqt, Koenigstein:2021llr} and \cref{sec:absence}). 
	Nevertheless, it is important to state that in the investigation of \gls{ff} models subjected to multiple, non-vanishing chemical potentials we restricted ourselves to a very limited set of interaction channels. 
	We want to highlight at this point that the restriction to a few interaction channels also limits the predictive power of our models for high energy phenomenology. For example, at finite isospin chemical potential one needs to take into account for charged pion condensation (see, e.g., \Rcite{Brandt:2017oyy}). Attentive readers may notice that the corresponding channels are not present in the models in \cref{tab:Results_chemipotentials}. In order to provide an adequate description of this phenomenon one certainly has to extend our study in this direction. 
	To study a larger set of interactions with several of the chemical potentials $\mu, \mu_{45},  \mu_I$ or even more axial imbalances\footnote{For example, one could introduce chemical potentials for the conserved currents $\bar{\psi} \gamma^\nu \gamma_{5} \psi$ or $\bar{\psi} \gamma^\nu \gamma_{4} \psi$ as in  \Rcite{ebert:2016ygm, zhukovsky:2017hzo}.} in models with more interaction channels, is in principle possible but requires a different, technically more involved analysis than the one done in this work.
	Thus, we postpone such an analysis to future works. However, our study allows us to conclude that the presence of multiple imbalances in fermion densities does not generically allow for the existence of inhomogeneous ground states.

	\begin{table*}[tbp]

		\begin{tabularx}{\textwidth}{>{\centering} p{2cm} >{\centering}  p{2cm} >{\centering}  p{2.5cm} >{\centering} p{3cm} >{\centering} p{4.5cm}  Y}
			\toprule
			\mycellcenter{Used \\ channels $\cm_j$}
			& \mycellcenter{Bosonic \\ auxiliary \\ fields $\auxf_j$} 
			& \mycellcenter{Non-zero \\ chemical \\ potentials}
			& Field basis $\vdbf_j$ diagonalizing $\seff^{(2)}$
			& \mycellcenter{Momentum dependence of $\gtwo_{\varphi_j}$ \\  \\ $f(M^2, \mu) = \EllTwoPlus(M^2, \mu, T, \qsV^2)$}  
			&  Underlying symmetry group  \\		
			\cline{1-6} \\ [-0.5em]
			$1, \gamma_{45}$
			& $\sigma, \eta_{45}$ 
			& \mycell{$\mu_L = \mu + \mu_{45}$ \\ $\mu_R = \mu - \mu_{45}$}
			& \mycell{$\phi_L = \left( \sigma + \eta_{45}\right)$\\  $\phi_R = \left( \sigma - \eta_{45}\right)$}
			& \mycell{$f(\Hom{\phi}_L^2, \mu_L)$ \\  $f(\Hom{\phi}_R^2, \mu_R)$}
			& $\hyperref[sym:U1_I]{\U_{\I_4}(N)} \times \hyperref[sym:U1_g45]{\U_{\gamma_{45}}(N)} \times \hyperref[sym:Z2_g5]{\Z_{\gamma_{5}}(2)} \times \hyperref[sym:U2_iso]{\SU_{\vpauli}(2)}\times \hyperref[sym:P4]{P_4} \times \hyperref[sym:P5]{P_5}$\\			
			\\ [-0.5em]	
			\cline{1-6}\\ [-0.5em]
			$1, \tau_3$
			& $\sigma, a_{0,3}$ 
			& \mycell{$\mu_\uparrow = \mu + \mu_{I}$\\ $\mu_\downarrow = \mu - \mu_{I}$}
			& \mycell{$\phi_\uparrow = \left( \sigma + \varsigma_{3}\right)$ \\ $\phi_\downarrow = \left( \sigma - \varsigma_{3}\right)$}
			& \mycell{$f(\Hom{\phi}_\uparrow^2, \mu_\uparrow)$ \\ $f(\Hom{\phi}_\downarrow^2, \mu_\downarrow)$}
			& $\hyperref[sym:U1_I]{\U_{\I_4}(N)} \times \hyperref[sym:U1_g45]{\U_{\gamma_{45}}(N)} \times \hyperref[sym:Z2_g5]{\Z_{\gamma_{5}}(2)} \times \hyperref[sym:U1_tau3]{\U_{\pauli_3}(1)}\times \hyperref[sym:P4]{P_4} \times \hyperref[sym:P5]{P_5}$ \\
			\\ [-0.5em]
			\cline{1-6} \\ [-0.5em]
			$1, \tau_3 \gamma_{45}$
			& $\sigma, \piveccomp{45}{3}$ 
			& \mycell{$\mu_{L, \uparrow} = \mu_L + \mu_{I}$ \\ $\mu_{L, \downarrow} = \mu_L - \mu_{I}$ \\ $\mu_{R, \uparrow} = \mu_R + \mu_{I}$\\ $\mu_{R, \downarrow} = \mu_R - \mu_{I}$}
			& \mycell{$\varphi_{+} = \left( \sigma + \piveccomp{45}{3}\right)$\\ \\ \\$\varphi_- = \left( \sigma - \piveccomp{45}{3}\right)$}
			&\mycell{$f(\Hom{\varphi}_+^2, \mu_{L, \uparrow})$ \\  $+ f(\Hom{\varphi}_+^2, \mu_{R, \downarrow}) $ \\  \\ $f(\Hom{\varphi}_-^2, \mu_{L, \downarrow})$ \\  $+ f(\Hom{\varphi}_-^2, \mu_{R, \uparrow}) $}
			& $\hyperref[sym:U1_I]{\U_{\I_4}(N)} \times \hyperref[sym:U1_g45]{\U_{\gamma_{45}}(N)} \times \hyperref[sym:Z2_g5]{\Z_{\gamma_{5}}(2)} \times \hyperref[sym:U1_tau3]{\U_{\pauli_3}(1)}\times \hyperref[sym:P4]{P_4} \times \hyperref[sym:P5]{P_5}$ \\
			\\ [-0.5em]	
			\cline{1-6}\\ [-0.5em]
			$1, \tau_3, \gamma_{45}$
			& $\sigma, a_{0,3}, \eta_{45}$
			& \mycellcenter{$\mu_{L, \uparrow}$, $\mu_{L, \downarrow} $, \\ $\mu_{R, \uparrow}$,  $\mu_{R, \downarrow} $}
			& \mycellcenter{$\phi_L = \left( \sigma + \eta_{45}\right)$\\ \\ \\  $\phi_R = \left( \sigma - \eta_{45}\right)$ \\ \\  \\$a_{0,3}$}
			& \mycellcenter{$f((\Hom{\phi}_L + \Hom{a}_{0,3})^2, \mu_{L, \uparrow})$ \\  $+ f(\Hom{\phi}_L - \Hom{a}_{0,3})^2, \mu_{L, \downarrow}) $ \\ \\ $f((\Hom{\phi}_R + \Hom{a}_{0,3})^2, \mu_{R, \uparrow})$ \\  $+ f(\Hom{\phi}_R - \Hom{a}_{0,3})^2, \mu_{R, \downarrow}) $ \\ \\ $\gtwo_{\phi_L} + \gtwo_{\phi_R}$}
			& $\hyperref[sym:U1_I]{\U_{\I_4}(N)} \times \hyperref[sym:U1_g45]{\U_{\gamma_{45}}(N)} \times \hyperref[sym:Z2_g5]{\Z_{\gamma_{5}}(2)} \times \hyperref[sym:U1_tau3]{\U_{\pauli_3}(1)}\times \hyperref[sym:P4]{P_4} \times \hyperref[sym:P5]{P_5}$\\		
			\\ [-0.5em]		
			\botrule
		\end{tabularx}

		\caption{Stability analysis of the bosonized \gls{ff} models with multiple chemical potentials and interaction channels. The first column describes the used \gls{ff} interaction vertices (compare \cref{eq:dirac_basis}) in the model. The second column gives the corresponding auxiliary bosonic fields after bosonization, that correspond to the fermion bilinear via the Ward identity \labelcref{eq:Ward}. The third column lists the used chemical potentials. In the fourth column, the field basis $\dbf_j$ is defined, which diagonalizes the second order correction, as described in detail in \cref{sec:res_mult_chempot}. Then, the momentum dependence of the two-point functions $\gtwo_{\dbf_j}$ is given in the fifth column, right to the field definition. The last column givens an overview of the present symmetries in the model. The groups are clickable and refer to the definition of the symmetry group.}

		\label{tab:Results_chemipotentials}
	\end{table*}

\subsection{Yukawa models \label{sec:res_Yukawa}}
	In this section, the results for the Yukawa model extension of the \gls{ff} models discussed above are presented. The generalization of the bosonized \gls{ff} models \labelcref{eq:part_bosonized} to a Yukawa model is discussed in \cref{sec:I_Yukawa}. In \cref{eq:Yukawa}, the Yukawa action is constructed out of the \gls{ff} models' effective action \labelcref{eq:Seff}. \\
	
	The homogeneous phase diagram of these models might drastically change compared to the \gls{ff} models due to the introduction of additional couplings and self-interaction terms. Nevertheless, the stability analysis, as described in \cref{sec:stability}, can be performed for all possible, homogeneous expansion points $\Yf_j = \HYf_j$ such that conclusions about the stability of these homogeneous condensates against inhomogeneous perturbations can be made. 
	
	First, we discuss the generalization of the results in \cref{sec:res_one_chempot}, where the \gls{ff} models are studied at finite baryon chemical potentials $\mu$. The key observation is that all bosonic two-point functions are positive, monotonically increasing functions of $\qsV^2$ and, thus, no instabilities are observed. 
	This observation is directly obtained for the Yukawa extensions of these models, as these only affect the momentum independent offset (compare \cref{eq:gtwoYukawaDiag}) and the physical expansion point $\minMass$.
	The momentum structure of the two-point function is still proportional to $\EllTwoPlusMinus$ plus an additional positive, monotonically increasing $\qsV^2$ term.
	As discussed in \cref{sec:absence} and illustrated by \cref{fig:ell2}, there is no expansion point for which $\EllTwoPlusMinus$ is not a monotonically rising function of $\qsV^2$.
	Thus, we can conclude that there exist no instabilities towards an \gls{ip} for the Yukawa models, which are generated by extending the \gls{ff} models in \cref{tab:Results}.
	According to the reasoning in \cref{sec:absence}, the moat regime with a negative wave-function renormalization is also ruled out.
	Nevertheless, it might be possible, that such a Yukawa model features a first order phase transition between the homogeneous phases.
	Then, one might also find the existence of a first order phase transition towards an \gls{ip}.
	In many model calculations the \gls{ip} covers the homogeneous first order transition \cite{ Thies:2003kk, Buballa:2014tba, Carignano:2014jla, Buballa:2018hux, Buballa:2020xaa} and, thus, the existence of such a transition could also mean that an \gls{ip} exists in these models, which is not detected by our analysis. 
	To our knowledge, however, a model, which features no instability towards an inhomogeneous perturbation in the whole phase diagram, but which still has an \gls{ip}, has never been observed before.
	
	Next, we discuss the extension of the \gls{ff} models from \cref{tab:Results_chemipotentials} with multiple chemical potentials to Yukawa models. 
	In the case of multiple chemical potentials, it is not straightforward to derive the two-point functions for the Yukawa models, corresponding to the \gls{ff} models by \cref{eq:Yukawa}, starting from the \gls{ff} model results.
	This is caused by the necessity of multiple non-vanishing homogeneous expectation values $\HYf_j$ as expansion points for the analysis when the models are subjected to multiple chemical potentials. 
	Then, the Yukawa self-interactions of the bosonic fields cause non-vanishing second order contributions, which are off-diagonal in the field perturbations $\ptYf_j$ (compare \cref{App:YukawaStab} and \cref{eq:sec_order_yukawa}), in addition to the off-diagonal fermion contributions \cref{eq:seff2_nondiag}. 
	This is the case for all Yukawa extensions of the models in \cref{tab:Results_chemipotentials}.
	
	In the analysis of such a model one needs to diagonalize the whole second-order contribution in \cref{eq:sec_order_yukawa}.
	Typically, we choose a basis $\dbYf_j$ for the dynamical scalar fields that corresponds to the field basis $\dbf_j$, which diagonalizes the \gls{ff} model part.
	Then, one still needs to diagonalize the off-diagonal contributions coming from the bosonic self-interactions.
	The diagonalization can become very complicated, even when using symbolic diagonalization of the expressions as provided by \texttt{Matlab} \cite{MATLAB}.
	We provide an example of this procedure in \cref{sec:yukawa_mul_cpot}, where we discuss the Yukawa extension of the model in the first row in \cref{tab:Results_chemipotentials}.
	In this analysis, we find a more complicated structure of the two-point functions and a diagonalizing field basis, which depends on the studied momentum $\qsV$ of the perturbation as well as the homogeneous expectation values of the scalar fields and the external parameters such as temperature and the chemical potentials. 
	Nevertheless, the obtained two-point functions can be shown to be monotonically increasing functions in $\qsV$ and, thus, again no instabilities towards inhomogeneous perturbations are observed. 
	For the other three models in \cref{tab:Results_chemipotentials}, we do not show an explicit calculation as the expressions become very lengthy and the analysis becomes very involved.
	However, by first calculations and due to the fact that the off-diagonal contributions coming from the scalar fields' self-interaction terms are not dependent on the momentum $\qsV$ of the inhomogeneous perturbations we expect that also these Yukawa models do not develop an instability towards an \gls{ip}.

\section{Conclusion and Outlook}
\label{sec:conclusion_and_outlook}

	In this work, we analyzed the stability of homogeneous condensates against inhomogeneous perturbations in a wide range of $2+1$-dimensional \glsfirst{ff} models and their Yukawa extensions under the influence of combinations of baryon chemical potential, isospin chemical potential and chiral chemical potential.
	All investigations were performed in the mean-field approximation, i.e., neglecting bosonic quantum fluctuations.
	The most involved model features $16$ Lorentz-(pseudo)scalar \gls{ff} interaction channels and the other models are subsets of this model (see \cref{eq:FFmodel,eq:dirac_basis} for the \gls{ff} model).
	
	Our main finding is the stability of homogeneous condensates against inhomogeneous perturbations in these \gls{ff} and Yukawa models at any finite baryon chemical potential and temperature.
	As argued in \cref{sec:absence}, this is strong evidence that \glspl{ip} do not exist in $2+1$-dimensional \gls{ff} models with Lorentz-(pseudo)scalar interaction channels and their corresponding Yukawa extension (the correspondence is given in \cref{sec:I_Yukawa}). 
	Also, we can completely rule out the existence of a moat regime whose characteristic is a negative wave-function renormalization $Z$ (see \cref{sec:intro} for a short discussion). 
	The momentum dependence of the obtained two-point functions only allow $Z\geq 0$.
	We draw similar conclusions for models with a subset of the above mentioned \gls{ff} interaction subjected to multiple chemical potentials.
	In the case of multiple chemical potentials, the extension to Yukawa models causes technical difficulties, but, in principle, we expect the result regarding stability of homogeneous phases to be the same. 
	This is motivated in \cref{sec:res_Yukawa} and an example computation for an explicit model is provided in \cref{sec:yukawa_mul_cpot}. 
	We note, however, that the used models are not very well suited for high energy phenomenology in the presence of multiple chemical potentials, as they lack, for example, the description of charged pion condensation for finite isospin chemical potential. Thus, one should interpret our results as a first step to generically show that multiple imbalances of fermion densities do not result in the existence of an \gls{ip}.

	Our results suggest that there might be a general argument, similar to Derrick's theorem \cite{Derrick:1964ww}, behind the absence of \glspl{ip} in $2+1$-dimensional models. 
	Such a principle could possibly be found by studying the properties of a general Ginzburg-Landau free energy as done in \Rcite{Pryadko_1999}. 
	Therefore, one would probably need to encode the fermionic determinant in our models by allowing for higher orders in the gradient expansion.  
	
	In the present study, we did not consider vector \gls{ff} interactions, since their inclusion yields a technically different analysis than the one presented in \cref{sec:stability}.
	The inclusion of these channels would also be relevant for low-energy effective theories of \gls{qcd}, where both vector and scalar \gls{ff} interaction should emerge. 
	However, the interplay of (repulsive) vector interactions and the scalar interactions could play a crucial role in developing an instability towards an \gls{ip}, especially as vector fields directly couple to fermionic momenta in the calculation\footnote{This is also the reason, why the analysis of these models differs from the ones presented in this work and has to be consistently presented elsewhere.}.
	In $3+1$-dimensional models it was already observed that vector interactions can enlarge the \gls{ip} \cite{Carignano:2010ac,Carignano:2018hvn}.
	The presence of such interactions causes additional contributions to appear in the two-point functions of the corresponding bosonic fields, which do not have a definite monotony as the momentum dependence of the two-point functions in the present work.
	Calculations for $2+1$-dimensional \gls{ff} models with vector interactions are already underway and we hope to report soon about the results.
	
	With respect to high energy phenomenology, there are many ways to build on the present work -- for example allowing for finite quark-masses, studying three-flavor versions of the \gls{ff} models allowing for strangeness effects or the inclusion of bosonic quantum fluctuations using lattice field theory.
	With respect to the stability analysis, it is interesting to note that using finite bare quark masses disfavors an \gls{ip} \cite{Buballa:2020xaa} and the inclusion of strange quarks only has marginal effects on inhomogeneous condensation \cite{Carignano:2019ivp}.
	However, the effect of bosonic quantum fluctuations beyond the mean-field approximation on the phase diagram is a very important aspect and also currently discussed in the literature \cite{Scherer:2012fjq, Lenz:2020bxk, Lenz:2021kzo, Stoll:2021ori,Lenz:2023wvk, Ciccone:2022zkg}. 
	The general result of these studies is that bosonic quantum fluctuations tend to disfavor ordered phases (such as condensed phases) similar to the effect of thermal fluctuations. Thus, it can be anticipated that the absence of an \gls{ip} in a mean-field calculation implies that there will be no inhomogeneous ground states in the full quantum theories. To our knowledge, there is no observation of an inhomogeneous ground state being generated by bosonic quantum fluctuations when there is no \gls{ip} present when these fluctuations are suppressed.
	An interesting aspect for $2+1$-dimensional theories in general could be the inclusion of difermion interactions in order to allow for color-superconducting order parameters, which often are in competition with the chiral ones, see, e.g., \Rcite{Sadzikowski:2006jq,ebert:2016ygm, Khunjua:2020xws, Lakaschus:2020caq}.

	Regarding the general understanding of strongly-interacting fermions in $2+1$ dimensions, it would be very interesting to study the interference and competition of the different chiral chemical potentials, which can be introduced as $\mu_{4} \bar{\psi} \gamma_3 \gamma_{4} \psi$, $\mu_{5} \bar{\psi} \gamma_3 \gamma_{5} \psi $ or $\mu_{45} \bar{\psi} \gamma_3 \gamma_{45} \psi $ (as done in the beginning of \cref{sec:res_mult_chempot}) and correspond to different conserved charges.
	However, the first results in our work and in \Rcite{Pannullo:2021edr} suggest that this would not change the phase diagram with respect to the existence of \glspl{ip}.

\begin{acknowledgments}
	Foremost, we want to thank Marc Wagner for supporting us and encouraging us to conduct this work.
	Moreover, we thank Adrian Koenigstein and Marc Wagner for very valuable comments on this manuscript.
	Furthermore, we acknowledge fruitful, related discussions with Michael Buballa, Adrian Koenigstein, Lennart Kurth, Julian Lenz, Michael Mandl, Zohar Nussinov,  Mike Ogilvie, Robert Pisarski,  Stella Schindler, Marc Wagner and Andreas Wipf.
	We acknowledge the support of the \textit{Deutsche Forschungsgemeinschaft} (DFG, German Research Foundation) through the collaborative research center trans-regio  CRC-TR 211 ``Strong-interaction matter under extreme conditions''-- project number 315477589 -- TRR 211.	
	We acknowledge the support of the \textit{Helmholtz Graduate School for Hadron and Ion Research}.	
	We acknowledge the support of the \textit{Giersch Foundation}.
	
\end{acknowledgments} 

\appendix
\section{Symmetries of \texorpdfstring{$(2 + 1)$}{(2+1)}-dimensional fermions\label{app:symmetries}}
	In this section, we want to elaborate on the symmetries of the fermionic models, which are introduced in \cref{sec:theory} in \cref{eq:FFmodel}. Their partially bosonized action can be found in \cref{eq:part_bosonized} and are invariant under certain symmetry transformations acting on the $2\N$ four-component spinor fields (the factor of $2$ comes from introducing an isospin space). The specific symmetry group, under which the model is invariant, is determined by the choice of the set $\indSet$, i.e.\ by the choice $\lambda_j \neq 0, \; j \in \indSet$ and, correspondingly, by which auxiliary bosonic fields $\auxf_j, \; j \in \indSet$ have to be introduced in the bosonization. This section is intended to provide an overview of the relevant symmetry groups present in the models discussed in \cref{sec:results} and the tables \cref{tab:Results} and \cref{tab:Results_chemipotentials}.
	
	Taking $\indSet = \{j\}_{j=1,\ldots,16}$ the $\N$ isospin up/down fermion fields in \cref{eq:part_bosonized} will be invariant under transformation of the group $\U(4\N)$, similar to the free case. This group has $\left(4 \N\right)^2$ generators, which we split up into three different categories. There are rotations within the internal space of the $\N$ spinors generated by $\N\times\N$ matrices given by the generalized Gell-Mann matrices $T^a$ with $a=1, \ldots, \N^2 - 1$ and the identity matrix $T^{\N^2} = \I_\N$. These generators $T^a$ can be combined with any of the chiral symmetry transformations of a free fermion fields, which are generated by $\left(\I_4, \gamma_{4}, \gamma_{5}, \gamma_{45}\right)$. Together with the  internal rotations either of these can generate a $\U(\N)$ symmetry, labeled by the corresponding chiral generator, namely  
	\begin{align}
		\U_{\I_4}(\N): \quad&\psi \rightarrow \mathrm{e}^{\ii\alpha^a\I_4 T^a}\psi\, ,\quad \bar{\psi} \rightarrow \bar{\psi} \mathrm{e}^{-\ii\alpha^a\I_4 T^a}\, , \label{sym:U1_I} \\
		\U_{\gamma_{45}}(\N): \quad&\psi \rightarrow \mathrm{e}^{\ii\beta^a\gamma_{45} T^a}\psi\, ,\quad \bar{\psi} \rightarrow \bar{\psi} \mathrm{e}^{-\ii\beta^a\gamma_{45} T^a}\,\label{sym:U1_g45}  , \\
		\U_{\gamma_{4}}(\N): \quad&\psi \rightarrow \mathrm{e}^{\ii\zeta^a\gamma_4 T^a}\psi\, ,\quad \bar{\psi} \rightarrow \bar{\psi} \mathrm{e}^{\ii\zeta^a\gamma_4 T^a}\,\label{sym:U1_g4} , \\
		\U_{\gamma_{5}}(\N): \quad&\psi \rightarrow \mathrm{e}^{\ii\iota^a\gamma_{5} T^a}\psi\, ,\quad \bar{\psi} \rightarrow \bar{\psi} \mathrm{e}^{\ii\iota^a\gamma_{5} T^a}\, , \label{sym:U1_g5}
	\end{align}
	with real parameters $\alpha^a, \beta^a, \zeta^a, \iota^a$. Thus, the whole chiral symmetry group can be defined as 
	\begin{equation}
		\U_{\gamma}(2\N): \quad \psi \rightarrow U \psi,  \label{sym:U2N_ch}
	\end{equation}
	where $U$ is a matrix element of $\U(2\N)$. 
	In addition, the FF model in \cref{eq:FFmodel} is invariant under isospin transformations of the group $\U(2)$, which is composed of a $\U(1)$ phase factor and  
	\begin{align}
		\SU_{\vpauli}(2):\quad &\psi \rightarrow \e^{\ii \ve{\xi} \vpauli} \psi\, ,\quad \bar{\psi} \rightarrow \bar{\psi} \e^{-\ii \ve{\xi} \vpauli}, \label{sym:U2_iso}
	\end{align}
	where $\ve{\xi}$ are three real parameters. \\
	
	Some of the models discussed in \cref{sec:results} are only invariant under a subgroup of the $\U(4\N)$ transformations. We will define these symmetry transformations as a reference for \cref{tab:Results} and \cref{tab:Results_chemipotentials}. As in the \gls{njl} model, one can choose $\indSet$ such, that one or more of the two axial transformations \labelcref{sym:U1_g4} and \labelcref{sym:U1_g5} are broken. Then, the action can be still invariant under a combined isospin and chiral rotation given by
	\begin{align}
		\SU_{A, \gamma_{4}}(2\N): \quad \psi \rightarrow \e^{\ii \ve{\zeta^\prime}^a \vpauli \gamma_{4} T^a  } \psi, \quad \bar{\psi} \rightarrow \bar{\psi} \e^{\ii \ve{\zeta^\prime}^a \vpauli \gamma_{4} T^a}, \label{sym:SU2_A_g4}\\
		\SU_{A, \gamma_{5}}(2\N): \quad \psi \rightarrow \e^{\ii \ve{\iota^\prime}^a \vpauli \gamma_{5} T^a  } \psi, \quad \bar{\psi} \rightarrow \bar{\psi} \e^{\ii \ve{\iota^\prime}^a \vpauli \gamma_{5} T^a}, \label{sym:SU2_A_g5}
	\end{align}
	where the isovectors $\ve{\iota^\prime}^a$ and $\ve{\zeta^\prime}^a$ contain $3$ real parameters and $a = 1, \ldots, \N^2$. Typically in these cases, the vector symmetry \cref{sym:U2_iso} is not broken.
	
	Some of the choices of $\indSet$ break the continuous chiral symmetries to discrete subgroups 
	\begin{align}
		\Z_{\gamma_{4}}(2): \psi \rightarrow \gamma_{4} \psi, \quad \bar{\psi} \rightarrow - \bar{\psi} \gamma_{4},\label{sym:Z2_g4} \\
		\Z_{\gamma_{5}}(2): \psi \rightarrow \gamma_{5} \psi, \quad \bar{\psi} \rightarrow - \bar{\psi} \gamma_{5}. \label{sym:Z2_g5}
	\end{align}
	Typically, when a model is in addition invariant under the transformation \labelcref{sym:U1_g45}, then either of \labelcref{sym:Z2_g4} and \labelcref{sym:Z2_g5} can be reproduced by a combination of \labelcref{sym:U1_g45} and the other of the discrete symmetries. 

	In \cref{tab:Results_chemipotentials}, only a remnant of the isospin symmetry transformation $\SU_{\vpauli}(2)$ is present in some of the models, namely
	\begin{equation}
		\U_{\pauli_3}(1):\quad \psi \rightarrow \e^{\ii \xi_3 \pauli_3} \psi\, ,\quad \bar{\psi} \rightarrow \bar{\psi} \e^{-\ii \xi_3 \pauli_3} \label{sym:U1_tau3}.
	\end{equation}

\section{Derivation of the stability analysis}
\label{app:stab}
	In this section, the stability analysis for a general FF theory subjected to a baryon chemical potential is derived.
	This discussion is similar to model-specific discussions, as, e.g., found in \Rcite{Buballa:2020nsi, Buballa:2020xaa, Koenigstein:2021llr}.
	The core idea of this technique is to analyze the stability of a homogeneous field configuration under inhomogeneous perturbations. 
	Furthermore, we outline how an extension to Yukawa models can be done.
	
	Consider an expansion of the auxiliary bosonic fields
	\begin{equation}
		\vauxf (\xsV) = \vHauxf  + \ptvauxf (\xsV),\label{eq:field_expansion2}
	\end{equation}
	where $\ptvauxf (\xsV)$ is a spatially dependent inhomogeneous perturbation of the homogeneous expansion point $\vHauxf$. 
	These perturbations are of an arbitrary shape and assumed to be of an infinitesimal amplitude.
	The Dirac operator then also separates into a homogeneous and an inhomogeneous part
	\begin{align}
		\D ={}&  \Big(\slashed{\partial}+ \gamma_3 \mu + \sum_{j\in \indSet} \, \cm_j \, \Hauxf_j\Big) +  \sum_k \cm_k \ptauxf_k(\xsV) \equiv \HD +   \ptD(\xsV),
	\end{align}
	which we can use to expand the $\ln \Det \D$ as
	\begin{align}
		&\ln \Det \left[\beta\D\right] =
	  \ln \Det \left[ \beta\HD\right] - \sum_{n=1}^{\infty} \tfrac{1}{n} \Tr\left[\left(- \ptD \HD^{-1} \right)^n\right],	\label{eq:lnDet_expansion}
	\end{align}
	where $\Tr$ denotes a functional trace over all spaces.
	We insert the expansion from \cref{eq:lnDet_expansion} and \cref{eq:field_expansion2} into the effective action \cref{eq:Seff} to obtain the expansion
	\begin{align}
		\frac{\seff[\vauxf]}{\N} = {}& \label{eq:expansion}  \int \dr^3x \,  \sum_{j\in \indSet} \frac{(\Hauxf  + \ptauxf(\xsV))^2 }{2 \coupling_j}  - \ln \Det \left[\beta\HD\right] + \sum_{n=1}^{\infty} \tfrac{1}{n} \Tr\left[\left(- \ptD \HD^{-1} \right)^n\right] \equiv   \frac{1}{N} \sum_{n=0}^{\infty} \seff^{(n)}, 
	\end{align}
	where $\seff^{(n)}$ contains all terms of order $n$ in the perturbations $\ptauxf_j$, i.e., terms $\propto \prod_{j} \ptauxf_j^{m_j}$ with $\sum_{j} m_j = n$.
	The first three terms in the series are given by
	\begin{align}
		\frac{\seff^{(0)}}{\N} ={}& \beta V \sum_{j\in \indSet} \frac{\Hauxf_j^2}{2 \coupling_j} - \ln \Det \left[\beta\HD\right]\ ,\label{eq:seff0}\\
		\frac{\seff^{(1)}}{\N} ={}& \beta \sum_{j\in \indSet} \frac{\Hauxf_j}{\coupling_j}\, \int \dr^2 x\, \ptauxf_j(\xsV) - \Tr\left[\ptD \HD^{-1}\,  \right]  ,\label{eq:seff1} \\
		\frac{\seff^{(2)}}{N} ={}&  \frac{\beta}{2} \sum_{j\in\indSet}  \frac{1}{\lambda_j} \int \dr^2 x\, \ptauxf_j^2(\xsV)  + \frac{1}{2} \Tr\left[\ptD \HD^{-1} \ptD \HD^{-1}  \right],
		\label{eq:seff2}
	\end{align}
	where the zeroth order term is proportional to the effective potential.
	In the position space representation, the fermion propagator depends only on the difference of two space-time variables, i.e., $\HD^{-1}=\HD^{-1}(\stV{x},\stV{y}) \equiv \HD^{-1}(\stV{x}-\stV{y})$. The functional traces are represented in position space as
	\begin{align}
		&\Tr\left[\left(\ptD\HD^{-1} \right)^n\right] = \int \prod_{j=1}^{n} \dr^3 x^{(j)} \tr\left( \ptD(\xsV^{(1)})\HD^{-1}\left(\xstV^{(1)}, \xstV^{(2)}\right) \ldots \ptD(\xsV^{(n)}) \HD^{-1}\left(\xstV^{(n)}, \xstV^{(1)}\right) \right),\nonumber
	\end{align}
	with $\tr$ denoting the trace in spinor and isospin space.
	In order to evaluate these traces, it is instructive to consider the Fourier representation of the homogeneous propagator
	\begin{align}
		&\HD^{-1}(\stV{x},\stV{y}) \equiv \label{eq:propagator_hom} \frac{1}{\beta} \matsubarasum \int \intMeasureOverPi{2}{p} \, \e^{\ii \left[\nu_n (\tau_x-\tau_y) + \sV{p} (\sV{x}-\sV{y})\right]} {\tilde{\HD}}^{-1}(\nu_n,\sV{p}), 
	\end{align}
	with
	\begin{align}
		{\tilde{\HD}}^{-1}(\nu_n,\sV{p}) ={}& \frac{-\ii \gamma_i \tilde{\stV{p}}_i + \sum_k \cmstar_k \Hauxf_k}{\tilde\nu_n^2 + \sV{p}^2+\Mass^2}
	\end{align}
	where $\tilde \nu_n = (\nu_n-\ii\mu)$, $\nu_n = 2 \uppi (n - \tfrac{1}{2}) / \beta$ are the fermionic Matsubara frequencies,  and $M$ and $\cmstar$ are defined in \cref{eq:MAndCStar}.
	Using the Fourier representation, \cref{eq:seff1} evaluates to
	\begin{align}
		\frac{\seff^{(1)}}{\N} ={}&   \sum_{j \in \indSet} \beta \Bigg[ \int \dr^2 \xstV \,  \ptauxf_j(\xsV)\Bigg] \label{eq:seff1Fourier}  \Bigg[ \frac{\Hauxf_j}{\coupling_j}    -  \frac{1}{\beta} \sum_n \int \intMeasureOverPi{2}{p}\tr(\ftHD^{-1}(\nu_n,\sV{p})\, c_j) \Bigg] , 
	\end{align}
	which we can identify to be proportional to the homogeneous gap equations 
	\begin{align}
		&\frac{1}{\N}\frac{\partial \seff^{(0)}}{\partial \Hauxf_k} \Bigg|_{\vHauxf = \vHauxf'} = \beta \Bigg[\int \dr^2 \xstV\Bigg] \Bigg[\frac{\Hauxf_k}{\coupling_k} - \frac{1}{\beta} \sum_n  \int \intMeasureOverPi{2}{p} \tr\left( \ftHD^{-1}(\nu_n, \sV{p})\, \cm_k \right) \Bigg]\Bigg|_{\vHauxf = \vHauxf'} = 0, \label{eq:gapequation}
	\end{align}
	where $\vHauxf'$ is a homogeneous field configuration that is a solution of the gap equations.
	Therefore, $\seff^{(1)}$ vanishes when the homogeneous field configurations used as an expansion point are solutions of the gap equations.
	If these expansion points are used, the second order term is the first non-zero correction. We evaluate the trace in \cref{eq:seff2} to
	\begin{align}
		&\Tr\left[ \HD^{-1}\, \ptD \HD^{-1}\, \ptD \right] = \label{eq:trace_ferm_loop} \int \intMeasureOverPi{2}{q}\, \left[\sum_{ j,k \in \indSet}\, \ptftauxf_j^*(\mathbf{q})\, \ptftauxf_k(\mathbf{q})\, \Gamma_{\phi_j \phi_k}^F \left(\Mass^2, \mu, T, q^2\right)\right], 
	\end{align}	
	where $q=|\qsV|$ and we used the Fourier representation of the spatial inhomogeneous perturbations
	\begin{align}
		\ptauxf_j(\xsV) = \int \intMeasureOverPi{2}{q} \e^{\ii \sV{q} \xsV}\, \ptftauxf_j(\qsV).
	\end{align}
	The matrix in field space $\Gamma^F_{\phi_j \phi_k}$ can be cast into the form
	\begin{align}	
		\Gamma^F_{\phi_j \phi_k} ={}& \frac{1}{\beta}\sum_n  \int \intMeasureOverPi{2}{p}\,  \mathrm{tr}\left(c_j\, \ftHD^{-1}(\nu_n,\sV{p} + \sV{q})\, c_k\, \ftHD^{-1}(\nu_n,\sV{p})\right) = \label{eq:one_loop_FF_secorder} \\
		={}&  -\frac{1}{\beta}\sum_n  \int \intMeasureOverPi{2}{p}\,  \frac{A_{c_j c_k}}{[\tilde \nu_n^2 + \sV{p}^2+\Mass^2][\tilde \nu_n^2 + (\sV{p}+\sV{q})^2+\Mass^2]} \nonumber 
	\end{align}
	with 
	\begin{align}
		A_{c_j c_k} ={}&  (\tilde \nu_n^2 + \sV{p}^2 + \sV{p} \cdot \qsV) \tr\left[\cm_j \gamma_i \cm_k \gamma_j\right]  \label{eq:l2_tr} -\sum_{l,m \in \indSet} \Hauxf_l \Hauxf_m \tr \Big[ \cm_j \cmstar_l \cm_k  \cmstar_m \Big] = \\
		={}& \delta_{j,k} \Nbar(\tilde \nu_n^2 + \sV{p}^2 + \sV{p} \cdot \qsV) -\!\!\!\!\sum_{l, m \in \indSet} \Hauxf_l \Hauxf_m \tr \Big[ \cm_j \cmstar_l \cm_k  \cmstar_m\Big], \nonumber
	\end{align}
	where we used that $\tr (\gamma_i \cm_j \cm_k \cm_l) = 0$, $\cm_j^2=\pm \I$ and that the anti-commutator $\left\{\gamma_i, \cm_k\right\}$ evaluates to $0$ or $2\cm_k\gamma_i$ for all considered $\cm_j$. 
	Thus, we obtain for the second order correction of the effective action
	\begin{align}
		\frac{\seff^{(2)}}{N} ={}& \frac{\beta}{2}  \int \intMeasureOverPi{2}{q}\, \sum_{j,k\in\indSet}\, \ptftauxf_j^*(\mathbf{q})\, \ptftauxf_k(\mathbf{q})\, \label{eq:seff2_nondiag} \left[\delta_{j,k}\,\lambda^{-1}_j+\Gamma_{\phi_j \phi_k}^F \left(\Mass^2, \mu, T, q^2\right) \right]. 
	\end{align}
	In order to make statements about the stability of a homogeneous field configuration one has to determine a basis $\dbf_j (\vauxf), \; j \in \indSet$ for which $\delta_{j,k}\,\lambda^{-1}_j+\Gamma_{\phi_j \phi_k}^F (q^2) $ is diagonalized. 
	This is not possible in general and depends on the present chemical potentials and the interactions of the model. 
	Furthermore, we assume that all $\lambda_j$ are either $\lambda$ or $0$ according to the subset $\indSet$ of the field content that we consider. 
	If this diagonalization is then indeed possible as it is the case for all channels considered in \cref{eq:FFmodel} at finite baryon chemical potential, one obtains the form
	\begin{align}
		\frac{\seff^{(2)}}{N} ={}& \frac{\beta}{2}  \int \intMeasureOverPi{2}{q} \left[\sum_{j \in \indSet}\, |\ptftdbf_j(\mathbf{q})|^2\, \Gamma_{\dbf_j}^{(2)}\left(\Mass^2, \mu, T, q^2\right) \right]  \label{eq:Seff_2_diag_app}
	\end{align}
	with
	\begin{widetext}
	\begin{align}
		&\Gamma_{\dbf_j}^{(2)}\left(\Mass^2, \mu, T, q^2\right) = \frac{1}{\coupling} -\frac{\Nbar}{\beta}\sum_n  \int \intMeasureOverPi{2}{p}\, \left(\frac{\tilde{\stV{p}}^2+ \sV{p}\cdot \sV{q}+a'_{\dbf_j} M^2 }{[\tilde \nu_n^2 + \sV{p}^2+\Mass^2][\tilde \nu_n^2 + (\sV{p}+\sV{q})^2+\Mass^2]}\right),  \label{eq:AppendixGamma2PrePartial}
	\end{align}
	\end{widetext}
	where $a'_{\dbf_j}$ is a coefficient that is determined by the considered field $\dbf_j$.
	In this diagonalized form, we identify $\Gamma_{\dbf_j}^{(2)}(q^2)$ as the curvature of the effective action for an inhomogeneous perturbation in field direction $\dbf_j$ with momentum $\sV{q}$. By writing the denominator of the integrand in \cref{eq:AppendixGamma2PrePartial} in a partial fraction, we can split the integral and obtain the final form of the bosonic two-point function as given in \cref{eq:gamma2}, where $ a_{\dbf_j}=2 (a'_{\dbf_j}-1)$. Note that the integral $\ell_1$ is also obtained in the fermionic trace in the gap equation \labelcref{eq:gapequation} for $\Hom {\phi}_1 = \Hom{\sigma}$ (compare also the \gls{gn} model gap equation in Section III of \Rcite{Buballa:2020nsi}). 
	The gap equation and, correspondingly, $\ell_1$ is typically used as a renormalization condition in the vacuum for the coupling constant $\coupling_j = \coupling$.

\subsection{The momentum independent part \texorpdfstring{$\EllOne$}{L1}}
\label{app:l1}
	We consider the integral
	\begin{align}
	\ell_1(\Mass^2,\mu,T)=	\frac{\Nbar}{\beta}\sum_n  \int \intMeasureOverPi{2}{p}\, \frac{ 1 }{(\nu_n - \ii \mu)^2+E^2},
	\end{align}
	where $E^2 = \sV{p}^2+\Mass^2$.
	The factor of $8$ comes from the traces over the four-dimensional spinor and the two-dimensional isospin space.
	Performing the sum over $n$, we obtain the standard result
	\begin{align}
		\ell_1(\Mass^2,\mu,T)= 	\Nbar \int \intMeasureOverPi{2}{p}\,	\frac{1-\ndist{E}{\mu} - \nbardist{E}{\mu}}{2 E},
	\end{align}
	where $n(x)$ is the fermionic distribution function
	\begin{align}
		n(x)=\frac{1}{\e^x+1}.
	\end{align}
	The vacuum part is UV-divergent, which we can regulate with a spatial momentum cutoff $\Lambda$
	\begin{align}
		\ell_1(\Mass^2,0,0)&= 	\frac{\Nbar}{2 \uppi} \int_0^\Lambda \dr p\,	\frac{p}{2 E} = \frac{\Nbar}{4 \uppi} (\sqrt{\Lambda^2 + \Mass^2} - |\Mass|) \stackrel{\Lambda \gg \Mass^2}{\to} \frac{\Nbar}{4 \uppi}  (\Lambda-|\Mass|)  \label{eq:ell1Vac}
	\end{align}
	and use to set the value of $\lambda_k$ via the gap-equations \cref{eq:gapequation}
	\begin{align}
		\frac{1}{\lambda_k} = \frac{\Nbar}{4 \uppi}  (\Lambda-|\minMass_0|) \label{eq:lambdaRen}
	\end{align}
	with the mass $\minMass_0$ corresponding to the minimum of the effective action in the vacuum (defined in \cref{sec:stability}).
	
	We are interested in the contribution of $\ell_1$ to the two-point function $\gtwo$, where $\ell_1$ appears exclusively as $\EllOne=\frac{1}{\lambda_k} - \ell_1$.
	Using \cref{eq:ell1Vac,eq:lambdaRen} we find 
	\begin{align}
		&\EllOneArgs{\Mass^2}{\mu}{T} = \Nbar\left[\frac{|\Mass|- |\Mass_0|}{4 \uppi}\right.+ \left. \int \intMeasureOverPi{2}{p}\,	\frac{\ndist{E}{\mu} + \nbardist{E}{\mu}}{2 E} \right],
	\end{align}
	where the medium integral over the fermionic-distribution function can be evaluated to 
	\begin{align}
		&\EllOneArgs{\Mass^2}{\mu}{T} = \frac{\Nbar}{4 \uppi} \Big[|\Mass|  - |\Mass_0| +\Big.\tfrac{1}{\beta}  \ln(1+ \e^{-\beta\left(|\Mass| - \mu\right)})+\tfrac{1}{\beta} \ln(1+ \e^{-\beta\left(|\Mass| + \mu\right)}) \Big]. 
	\end{align}
	For $T=0$, this evaluates to 
	\begin{align}
		&\EllOneArgs{\Mass^2}{\mu}{T=0} 
		= \frac{\Nbar}{4 \uppi}\left[|\Mass| - |\Mass_0| + \Theta\left(\mu^2 - \Mass^2\right) (|\mu| - |\Mass|)  \right],
	\end{align}
	from which we can naively take the limits $\mu \to 0$ and/or $\Mass \to 0$.

\subsection{The momentum dependent part \texorpdfstring{$\EllTwo{\pm}$}{L2}}
\label{app:l2}
	In order to calculate the momentum dependent part of the two-point function, we start by carrying out the Matsubara summation in $\ell_2$ and obtain
	\begin{widetext}
	\begin{align}
		\ell_2(\Mass^2,\mu,T,
		q^2)=\Nbar \int \intMeasureOverPi{2}{p} \frac{1}{2 \sV{p} \cdot \sV{q} + q^2} \left[ \frac{ 1 - \ndist{E}{\mu}- \nbardist{E}{\mu} }{ 2E}  -  \frac{ 1 - \ndist{E_\sV{q}}{\mu}- \nbardist{E_\sV{q}}{\mu}  }{ 2E_\sV{q}} \right],
	\end{align}
	\end{widetext}
	where $E_\sV{q} = \sqrt{\Mass^2+(\sV{p+q})^2}$. 
	This integral is UV-finite and, thus, we do not have to implement a regularization scheme. However, the integrand has a divergence at $2 \sV{p} \cdot \sV{q} = -q^2$ that has to be treated with a Cauchy principal value prescription.
	The vacuum contribution can be calculated analytically and we obtain
	\begin{align}
	&\ell_2(\Mass^2,\mu,T,
		q^2)=\frac{\Nbar}{4 \uppi q} \Bigg[ \arctan\left(\frac{q}{2 |\Mass|}\right)- \int_0^{q/2} \dr p \, \frac{p}{E} \frac{ \ndist{E}{\mu}+ \nbardist{E}{\mu}}{\sqrt{q^2/4-p^2}}  \Bigg].
	\end{align}
	At any finite $T$, the medium contribution has to be calculated numerically. 
	However, taking the limit $T\to0$ enables us to also calculate the medium contribution analytically and we find \cref{eq:ell2T0}.
	From this expression, we can take either the limit $q \to 0$ to obtain
	\begin{align}
		\ell_2(\Mass^2,\mu,T=0,q^2=0)&= \frac{1 }{\uppi }
		\begin{cases} 
			0\,,\quad &\mu^2 > \Mass^2  \vphantom{\Bigg[]}\\
			\frac{1}{|\Mass|} \,,\quad &\mu^2 < \Mass^2 \vphantom{\Bigg[]}\\					
		\end{cases}	
	\end{align}
	or the  limit $|\Mass| \to 0$ to obtain
	\begin{align}
		&\ell_2(\Mass^2=0, \mu, T=0, q^2)=\frac{2}{\uppi q}
		\begin{cases} 
			0 \,,\quad &\mu^2 >  q^2/4  \\
			\arctan \left(\frac{\sqrt{q^2 -4\mu^2)}}{2\mu}\right)\,,\ & 0< \mu^2 \leq  q^2/4\   \nonumber\\	
			\frac{\uppi}{2} \,,\quad &\mu^2 =0
		\end{cases}	. 
	\end{align}
	While $\ell_2$ is not defined for $\Mass=q=T=0$ for some values of $\mu$, the whole momentum-dependent contribution to the two-point functions $\EllTwo{\pm}$ is defined with
	\begin{align}
		\EllTwoArgs{\pm}{\Mass^2=0}{\mu}{T=0}{q^2=0}=0.
	\end{align}

\subsection{Generalization to Yukawa models\label{App:YukawaStab}}
	It is straightforward to generalize the stability analysis of \gls{ff} models to Yukawa models, which are defined as in \cref{sec:I_Yukawa} in \cref{eq:Yukawa}. Thus, we outline only the meaningful differences with respect to the discussion of \gls{ff} models in order to preserve brevity.
	
	After an expansion of $\S_{Y}$ in powers of inhomogeneous perturbations of the fields, one identifies again $\S_{Y}^{(0)}$ as proportional to the effective potential. The first order correction $\S_{Y}^{(1)}$ is proportional to the gap equations and, consequently, vanishes when one expands about homogeneous extrema of $\S_{Y}$.
	
	For the second order correction $\S_{Y}^{(2)}$ we find
	\begin{widetext}
		\begin{align}
			\frac{\S_{Y}^{(2)}}{N} ={}&  \frac{\beta}{2} \sum_{ j \in \indSet}  \frac{1}{h \lambda_j} \int \dr^2 x\, \ptYf_j^2(\xsV)   - \Tr\left[ \HD^{-1}\,h \ptD \HD^{-1}\, h\ptD \right] +\label{eq:sec_order_yukawa}\\
			&+ \frac{1}{2}  \sum_{ j \in \indSet} \left\{ \int \dr^2 x\, \left(\partial_\nu \ptYf_j(\xsV)\right) \left(\partial_\nu \ptYf_j(\xsV)\right) + \sum_{n > 1}\kappa_n 2n\left(\vHYf^2\right)^{n-1}  \int \dr^2 x\, \ptYf_j^2(\xsV)\right\} + \nonumber\\
			&+ 2\sum_{ j,k \in \indSet } \sum_{n>1} \kappa_n n(n-1)  \HYf_j \HYf_k \left(\vHYf^2\right)^{n-2} \int \dr^2 x\, \ptYf_j(\xsV) \ptYf_k(\xsV),\nonumber
		\end{align}
	\end{widetext}
	where the second line and third line contain the additional terms resulting from the extension to Yukawa models. Note that the third line contains non-diagonal contributions from the self-interaction terms of the $\vYf$ fields. However, these are proportional to $\HYf_j \HYf_k$, and, thus, vanish when either of $\HYf_j, \HYf_k$ can be rotated to zero through a symmetry transformation, as is the case for all Yukawa-type extensions of the model \cref{eq:Seff}, where only a baryon chemical potential is present.
	If this is not the case, one needs to take into account this off-diagonal contribution in addition to the, in principle, off-diagonal fermionic contribution.
	This is the case for the models discussed in \cref{tab:Results_chemipotentials} with additional chemical potentials.
	A more involved diagonalization needs to be performed, although the Yukawa contribution is not expected to change the general momentum-dependence of the two-point function, as the term is $\qsV$-independent.
	An example for such an analysis is presented in \cref{sec:yukawa_mul_cpot}.
	
	In the case of only a baryon chemical potential, we utilize symmetry transformations to obtain a homogeneous expansion point $\vYf = \vHYf$ such that the off-diagonal contributions in the third row of \cref{eq:sec_order_yukawa} vanish. Performing similar steps as between \cref{eq:l2_tr} and \cref{eq:gamma2} leads to the second order correction \cref{eq:seff2Yukawa} and the corresponding bosonic two-point function \cref{eq:gtwoYukawaDiag}.

\section{Stability analysis for a Yukawa model with more than one chemical potential \label{sec:yukawa_mul_cpot}}
	In this section, we will show how the off-diagonal contribution from the Yukawa self-interactions (compare the third line of \cref{eq:sec_order_yukawa}) makes the diagonalization of $\S_{Y}^{(2)}$ more involved. 
	However, we will also demonstrate that this $\qsV$-independent contribution does not alter the predictions coming out of the analysis. 
	
	The model that we will study is defined as the Yukawa model extension according to \cref{eq:Yukawa} of the \gls{ff} model in the first row of \cref{tab:Results_chemipotentials}, i.e., it contains the $\sigma$ and $\eta_{45}$ fields as well as a baryon chemical potential $\mu$ and a chiral chemical potential $\mu_{45}$. 
	As documented in in \cref{tab:Results_chemipotentials} and \cite{Winstel:2022jkk}, the \gls{ff} part of the model is diagonalized by the field basis proportional to
	\begin{equation}
	 \left( \sigma \pm \eta_{45}\right).
	\end{equation}
	In analogy to this \gls{ff} model, we study the effective action 
	\begin{align}
		\frac{\seff[\chi_L, \chi_R]}{\N}  ={}& \label{eq:yukawa_example}  - \Tr \ln \left[\slashed{\partial}  +\right.  \left. \gamma_3\left(P_L \mu_L + P_R \mu_R \right)  + h P_L \chi_L  + h P_R \chi_R   \right] + \\
		&+ \int \dr^{3} x \, h^2\Big[  \tfrac{\chi_L^2 + \chi_R^2}{2 \coupling} + \tfrac{1}{2} \left(\partial \chi_L\right)^2 + \tfrac{1}{2} \left(\partial \chi_R\right)^2  +    \sum_{n> 1} \kappa_n h^{2(n-1)} \left( \chi_L^2 + \chi_R^2\right)^n \Big] \nonumber,
	\end{align}
	where $\chi_{L/R}$ are fields of canonical dimension and proportional to the dynamical scalar fields $\chi_\sigma$ and $\chi_{\eta_{45}}$ as 
	\begin{equation}
		\chi_L = \tfrac{1}{\sqrt{2}}\left( \chi_\sigma + \chi_{\eta_{45}}\right), \quad \chi_R = \tfrac{1}{\sqrt{2}}\left( \chi_\sigma - \chi_{\eta_{45}}\right).
	\end{equation}
	We, again, introduced a Yukawa coupling $h$ as well as couplings $\kappa_n$ for the self-interactions and define projectors and chemical potentials accordingly
	\begin{align}
		P_L = \tfrac{1}{\sqrt{2}} \left(1 + \gamma_{45} \right), \quad P_R = \tfrac{1}{\sqrt{2}} \left(1 - \gamma_{45} \right), \\
		\mu_L = \tfrac{1}{\sqrt{2}}\left(\mu + \mu_{45}\right), \quad \mu_R = \tfrac{1}{\sqrt{2}} \left(\mu - \mu_{45}\right).
	\end{align} 
	All terms except for the last row in \cref{eq:yukawa_example} either contain only\footnote{The Dirac operator within the $\Tr\ln$ can be decomposed into a block-diagonal form, where each block only contains either $\mu_L$ and $\chi_{L}$ or $\mu_R$ and $\chi_{R}$. In this sense, the fermionic contributions completely decouple $\chi_{L}$ and $\chi_{R}$.} $\chi_L$ or only $\chi_R$. Thus, the second order correction is given by 
	\begin{align}
	\label{eq:seff2_yuk_example}	\frac{\seff^{(2)}}{\N} ={} \frac{\beta}{2} \int \tfrac{\dr^2 q}{(2\uppi)^2}  \Bigg\{ &\sum_{j=L,R} |\pt{\chi}_j(\sV{q})|^2 \Bigg[ h^2\Gamma^{F}_{\chi_j} + q^2  + \sum_{n>1} \kappa_n n \left( 2(\Mass^2)^{n-1} + 4 (n-1) \bar{\chi}_j^2 (\Mass^2)^{n-2} \right) \Bigg] \Bigg. +\\ & \Bigg. + \pt{\chi}_L (-\qsV) \pt{\chi}_R(\qsV) 4 \kappa_n n (n-1) \bar{\chi}_L \bar{\chi}_R (\Mass^2)^{n-2} + L \leftrightarrow R \Bigg\} \nonumber,  
	\end{align}
	where $\Mass^2 = \bar{\chi}_\sigma^2 + \bar{\chi}_{\eta_{45}}^2 = \bar{\chi}_L^2 + \bar{\chi}_{R}^2 $ and 
	\begin{equation}
		\Gamma^{F}_{\chi_j}  = \frac{1}{\coupling} - \ell_1 + L_{2, +}(h^2\bar{\chi}_j^2, \mu_j, T, q^2)
	\end{equation}
	is the contribution, that also appears in the corresponding \gls{ff} model (see \cref{tab:Results_chemipotentials} and \Rcite{Winstel:2022jkk}). 
	The integrals $\ell_1$ and  $L_{2, +}$ are defined in \cref{eq:gamma2,eq:L2pm}.
	
	The last row of \cref{eq:seff2_yuk_example} contains the off-diagonal contribution of the self-interaction. 
	This contribution is not dependent on the spatial momentum $\mathbf{q}$ of the perturbation, but it makes the diagonalization more complicated (compare \cref{eq:sec_order_yukawa} for the form of this contribution in the more general case). 
	In fact, we are only able to diagonalize this symbolically using \texttt{Matlab} \cite{MATLAB}. Using the definitions 
	 \begin{equation}
	 	Y_j =  q^2 + \sum_{n>1} \kappa_n n \left( 2(\Mass^2)^{n-1} + 4 (n-1) \bar{\chi}_j^2 (\Mass^2)^{n-2} \right),
	 \end{equation}
	 with $j = L, R$ as well as 
	 \begin{equation}
	 	I = \sum_{n>1} 4 \kappa_n n (n-1) \bar{\chi}_L \bar{\chi}_R (\Mass^2)^{n-2}
	 \end{equation}
	 and 
	 \begin{equation}
	 	A = \sqrt{(h^2\Gamma^F_{\chi_{L}} - h^2\Gamma^F_{\chi_{R}} + Y_L - Y_R)^2 +  (2I)^2}
	 \end{equation}
	 we write the diagonalization of $\seff^{(2)}$ in the $\left(\pt{\chi}_L (\qsV), \pt{\chi}_R(\qsV)\right)$-space 
	 \begin{widetext}
	 \begin{align}
	 	\label{eq:seff2_yuk_example_diag} \frac{\seff^{(2)}}{\N} ={} & \frac{\beta}{2} \int \tfrac{\dr^2 q}{(2\uppi)^2} \left(\pt{\chi}_L (-\qsV), \pt{\chi}_R(-\qsV) \right) B(q^2, \bar{\chi}_L, \bar{\chi}_R, \mu_L, \mu_R, T) \times \\ & \nonumber
	 	 \times \begin{pmatrix}
	 	 \tfrac{h^2}{2} \left(\Gamma^F_{\chi_{L}} + \Gamma^F_{\chi_{R}} \right) + \tfrac{1}{2} \left(Y_L + Y_R\right) - \tfrac{1}{2} A &  \\
	 	  & \tfrac{h^2}{2} \left(\Gamma^F_{\chi_{L}} + \Gamma^F_{\chi_{R}} \right) + \tfrac{1}{2} \left(Y_L + Y_R\right) + \tfrac{1}{2} A 
	 	 \end{pmatrix} \times \\
	  & \times  B^{-1} (q^2, \bar{\chi}_L, \bar{\chi}_R, \mu_L, \mu_R, T) \left(\pt{\chi}_L (\qsV), \pt{\chi}_R(\qsV) \right)^T, \nonumber
	 \end{align}
	\end{widetext} 
	where $B(q^2, \bar{\chi}_L, \bar{\chi}_R, \mu_L, \mu_R, T)$ is a basis changing matrix determined by \texttt{Matlab}, whose form is not relevant for our analysis. 
	In this form one can determine, whether the diagonal entries of the matrix in \cref{eq:seff2_yuk_example_diag} are non-negative.
	For the physically relevant homogeneous expansion point $\minMass$ both entries are non-negative for $\qsV = 0$, since otherwise the expansion point would not be a minimum when only considering homogeneous field values.
	Therefore, in order to prove positivity for all $q=|\qsV|$ it suffices again to show that the entries are monotonically increasing functions of $q$.
	We take the derivative of the entries with respect to $q$ and require it to be non-negative
	\begin{widetext}
	\begin{align}
		\frac{h^2}{2}\left[\left(L_{2,L} \right)'+\left(L_{2,R} \right)'\right] + 2 q \mp \frac{h^2}{2} \left[\left(L_{2,L} \right)'-\left(L_{2,R} \right)'\right] \frac{h^2\Gamma^F_{\chi_{L}} - h^2\Gamma^F_{\chi_{R}} + Y_L - Y_R }{A} \stackrel{!}{\geq} 0, \label{eq:yuk_example_entry_derivative}
	\end{align}
	where $L_{2,L/R}=\EllTwoArgs{+}{h^2\bar{\chi}_{L/R}^2}{\mu_{L/R}}{T}{ q^2}$ and its derivative with respect to $q$ is non-negative, i.e., $ \tfrac{\dr }{\dr q}L_{2,L/R} =\left(L_{2,L/R} \right)'\geq 0$, since it is a monotonically increasing function of $q$ (compare \cref{sec:stability}).
	We can rearrange \cref{eq:yuk_example_entry_derivative} and square it to obtain
	\begin{align}
		\left(\left[\left(L_{2,L} \right)'+\left(L_{2,R} \right)'\right] + \frac{2 q}{h^2}\right)^2   \geq {}& \left[\left(L_{2,L} \right)'-\left(L_{2,R} \right)'\right]^2 \frac{\left(h^2\Gamma^F_{\chi_{L}} - h^2\Gamma^F_{\chi_{R}} + Y_L - Y_R \right)^2}{(h^2\Gamma^F_{\chi_{L}} - h^2\Gamma^F_{\chi_{R}} + Y_L - Y_R)^2 +  (2I)^2} = \\ ={}& \left[\left(L_{2,L} \right)'-\left(L_{2,R} \right)'\right]^2 c^2 , \nonumber
	\end{align}
	where obviously $0 \leq c^2 \leq 1$ and, thus, the inequality is fulfilled for all $q$.
	\end{widetext}
	
	Summarizing this lengthy and delicate analysis: We diagonalized the second order corrections \labelcref{eq:seff2_yuk_example} of a Yukawa model with multiple chemical potentials given by \cref{eq:yukawa_example} using computer algebra systems such as \texttt{Matlab} \cite{MATLAB}. 
	Analyzing the resulting expression, we find that both eigenvalues of the relevant curvature matrix in the second order corrections are positive, monotonically increasing functions of the momentum squared $q^2$ of the inhomogeneous perturbation. Thus, we do not observe instabilities of the homogeneous condensates in the Yukawa model given by \cref{eq:yukawa_example}.
	By the same reasoning, a negative wave-function renormalization (proportional to the second derivative of the two eigenvalues with respect to $q$), i.e., a so-called moat regime is not observed in the model.
	Similar behavior is expected for the other Yukawa models that correspond to the \gls{ff} models in \cref{tab:Results_chemipotentials} according to \cref{sec:I_Yukawa,eq:Yukawa}.

\bibliography{literature}

\end{document}